\documentclass[12pt,preprint]{aastex}

\shorttitle{THE EFFICIENCY AND WAVELENGTH DEPENDENCE OF NIR INTERSTELLAR POLARIZATION TOWARD THE GC}
\shortauthors{Hatano et al.}

\begin{document}

\title{THE EFFICIENCY AND WAVELENGTH DEPENDENCE OF NEAR-INFRARED INTERSTELLAR POLARIZATION TOWARD THE GALACTIC CENTER}

\author{Hirofumi Hatano,\altaffilmark{1} Shogo Nishiyama,\altaffilmark{2} Mikio Kurita,\altaffilmark{1}$^,$\altaffilmark{3}
Saori Kanai,\altaffilmark{1} Yasushi Nakajima,\altaffilmark{2}$^,$\altaffilmark{4} Tetsuya Nagata,\altaffilmark{3}
Motohide Tamura,\altaffilmark{2} Ryo Kandori,\altaffilmark{2} Daisuke Kato,\altaffilmark{5} Yaeko Sato,\altaffilmark{6}
Tatsuhito Yoshikawa,\altaffilmark{3} Takuya Suenaga,\altaffilmark{6} and Shuji Sato\altaffilmark{1}}

\altaffiltext{1}{Department of Astrophysics, Nagoya University, Chikusa-ku, Nagoya 464-8602, Japan;
                 hattan@z.phys.nagoya-u.ac.jp.}
\altaffiltext{2}{National Astronomical Observatory of Japan, Mitaka, Tokyo 181-8858, Japan;
                 shogo.nishiyama@nao.ac.jp.}
\altaffiltext{3}{Department of Astronomy, Kyoto University, Sakyo-ku, Kyoto 606-8502, Japan.}
\altaffiltext{4}{Center for Information and Communication Technology, Hitotsubashi University, Kunitachi, Tokyo 186-8601, Japan.}
\altaffiltext{5}{Department of Astronomy, School of Science, University of Tokyo, Bunkyo-ku, Tokyo 113-0033, Japan.}
\altaffiltext{6}{Department of Astronomical Sciences, Graduate University for Advanced Studies (Sokendai),
                 Mitaka, Tokyo 181-8858, Japan.}

\begin{abstract}

Near-infrared polarimetric imaging observations toward the Galactic center have been carried out to examine
the efficiency and wavelength dependence of interstellar polarization.
A total area of about 5.7 deg$^2$ is covered in the $J$, $H$, and $K_S$ bands.
We examined the polarization efficiency, defined as the ratio of degree of polarization to color excess.
The interstellar medium between the Galactic center and us shows the polarization efficiency lower than that in the Galactic disk
by a factor of three.
Moreover we investigated the spatial variation of the polarization efficiency
by comparing it with those of color excess, degree of polarization, and position angle.
The spatial variations of color excess and degree of polarization depend on the Galactic latitude,
while the polarization efficiency varies independently of the Galactic structure.
Position angles are nearly parallel to the Galactic plane,
indicating the longitudinal magnetic field configuration between the Galactic center and us.
The polarization efficiency anticorrelates with dispersions of position angles.
The low polarization efficiency and its spatial variation can be explained by
the differences of the magnetic field directions along the line-of-sight.
From the lower polarization efficiency, we suggest a higher strength of a random component
relative to a uniform component of the magnetic field between the Galactic center and us.
We also derived the ratios of degree of polarization $p_H/p_J$ = 0.581 $\pm$ 0.004 and $p_{K_S}/p_H$ = 0.620 $\pm$ 0.002.
The power law indices of the wavelength dependence of polarization are
$\beta_{JH}$ = 2.08 $\pm$ 0.02 and $\beta_{HK_S}$ = 1.76 $\pm$ 0.01.
Therefore the wavelength dependence of interstellar polarization exhibits flattening
toward longer wavelengths in the range of 1.25$-$2.14 $\micron$.
The flattening would be caused by aligned large-size dust grains.

\end{abstract}

\keywords{polarization --- dust, extinction --- ISM: magnetic fields --- Galaxy: center --- infrared: stars}

\section{INTRODUCTION}

Observations of interstellar linear polarization (hereafter just interstellar polarization)
give information about the magnetic field and properties of polarizing dust grains.
Interstellar polarization is caused by non-spherical dust grains aligned by a magnetic field
\citep[dichroic extinction; see e.g., reviews by][]{Lazarian03,Lazarian07}.

Position angles of polarization yield the directions of magnetic fields on the plane-of-the-sky.
\citet{Mathewson70} compiled the polarization data of nearly 7000 stars and showed the distribution of position angles
in the Galactic coordinates.
The distribution of position angles for stars beyond 1 kpc of the Sun traces the structure of the large-scale Galactic magnetic field
which runs almost parallel to the spiral arms, while that for stars within 600 pc traces the structure of the local magnetic field.
The structure of magnetic field is less simple at $l$ $\sim$ 40$\arcdeg$, 80$\arcdeg$, 260$\arcdeg$,
and away from the Galactic plane such as the active star forming regions Taurus, Perseus, Ophiuchus, and Orion,
where position angles deviate from the large-scale longitudinal pattern.
The local magnetic field points toward $l$ $\sim$ 80$\arcdeg$ and away from $l$ $\sim$ 260$\arcdeg$ \citep{Heiles76}.
From the direction of $l$ $\sim$ 40$\arcdeg$, a loop structure extends toward the north Galactic pole.
Toward the loop structure and the star forming regions,
the magnetic field is obviously perturbed from the uniform, large-scale structure.
These indicate the existence of a random component of the magnetic field on a small-scale
in addition to a uniform component on a large-scale.

The random component of the magnetic field can cause a dispersion in the measured position angles
and a decrease of the polarization efficiency.
The polarization efficiency is defined as a ratio of polarization degree to extinction such as
$p_\lambda/\tau_\lambda$ and $p_{\lambda_2}/E(\lambda_1 - \lambda_2)$.
If magnetic fields are tangled along the line-of-sight, degree of polarization $p_\lambda$ does not build up
as much as it would in a uniform magnetic field.
The extinction (as measured by $\tau_\lambda$ or $E(\lambda_1 - \lambda_2)$) along a line-of-sight is determined only by
the total column of dust, so that the polarization efficiency decreases \citep[depolarization;][]{Martin74}.
Thus, the polarization efficiency is a useful measure to probe the random component of the magnetic field.

\citet{Serkowski75} analyzed a sample of 180 nearby stars that were observed polarimetrically in the $U$, $B$, $V$, and $R$ bands.
They investigated the relation between color excess $E(B - V)$ and the maximum polarization $p_\mathrm{max}$
at the wavelength $\lambda_\mathrm{max}$, and found an upper limit for the polarization efficiency,
\begin{displaymath}
  p_\mathrm{max}/E(B-V) \leq 9.0\% / \mathrm{mag}.
\end{displaymath}
The polarization efficiency changes from line-of-sight to line-of-sight below the upper limit.

Observations at optical wavelengths are limited to regions with small extinction (typically $A_V$ $<$ 5 mag).
The behavior of the polarization efficiency at large extinction (up to $A_V$ $\sim$ 100 mag)
was studied by \citet{Jones89} and \citet{Jones92} using near-infrared (NIR) wavelengths.
They compiled $K$ band (2.2 $\micron$) polarimetric measurements for about 100 sources at various locations and extinctions.
The relation between polarization and extinction was modeled by assuming that interstellar polarization depends only on
the geometry of the uniform and random components of the magnetic field.
The model fits the observed relation between polarization and extinction well
when equipartition of the energy density holds between the uniform and random components of the magnetic field.

Studies on the polarization efficiency of the diffuse interstellar medium (ISM) have been carried out
using the comprehensive compilation of polarization data of \citet{Heiles00}.
Based on the polarization data for about 5,500 stars distributed over the entire sky
and mostly located at distances of $d$ $\lesssim$ 4 kpc, \citet{Fosalba02} found
a nearly linear growth of average polarization degree with extinction up to $E(B - V)$ $\sim$ 1 mag, but noted that
the polarization efficiency is much lower than what is expected from completely aligned grains in a uniform magnetic field.
They found the polarization efficiency that was 1/3 that of the maximum found by \citet{Serkowski75}.
They explained it by depolarization due to the random component of the magnetic field,
and estimated the magnetic field strength ratio of the uniform to the random component, $B_u/B_r$, to be about 0.8,
where $B_u$ and $B_r$ are the strengths of the uniform and random components.

In the innermost region of the Galaxy, only a few polarimetric observations have been done
\citep[e.g.,][]{Kobayashi83,Kobayashi86,Creese95}.
\citet{Kobayashi83,Kobayashi86} measured polarization for a few dozen of highly reddened ($H - K$ $\lesssim$ 3 mag) stars
in the area at $l$ $\sim$ 0$\arcdeg$, 20$\arcdeg$, and 30$\arcdeg$ in the Galactic plane,
and noted that the polarization efficiency is lower by a factor of about four than that in the solar neighborhood.
Based on polarization measurements for 127 reddened stars, \citet{Creese95} reported that
greater extinction results in increased polarization, but the increase is smaller than expected.
They concluded that the polarization efficiency is lower than that in the solar neighborhood.
Their samples are not highly extincted ($H - K$ $\lesssim$ 0.7 mag),
because they were selected from an $I$ band objective prism survey and were therefore relatively sparse and shallow.

The properties of polarizing dust grains can be examined by the wavelength dependence of interstellar polarization.
Polarization $p_\lambda$ shows a convex curvature with a peak $p_\mathrm{max}$, typically occurring at around
$\lambda_\mathrm{max}$ = 0.55 $\micron$, and a wing toward NIR wavelengths.
\citet{Serkowski75} made a determination of the wavelength dependence of polarization, empirically establishing ``Serkowski's law'',
\begin{displaymath}
  p_\lambda/p_\mathrm{max} = \mathrm{exp} [-1.15\mathrm{ln^2} (\lambda_\mathrm{max} / \lambda)].
\end{displaymath}
The slope of the wing is represented by a power law as
\begin{displaymath}
  p_\lambda \propto \lambda^{-\beta}
\end{displaymath}
with $\beta$ of 1.6$-$2.0 from 1.25 to 2.2 $\micron$ \citep{Nagata90,Martin90,Martin92}.
\citet{Creese95} suggest no systematic trend in the observed $JHK$ polarization for about 10 reddened stars
with respect to a power law.
However, more samples should be needed to confirm this suggestion.

Now polarimetric imaging observations using several hundred thousand stars lying in the Galactic bulge as background sources
enable us to measure polarization degrees, position angles, and color excess finely across the area and deeply to the line-of-sight.
Therefore, we carried out NIR polarimetric imaging observations for a wide field ($\sim$ 3$\arcdeg$ $\times$ 2$\arcdeg$)
toward the Galactic center (GC).
From the polarization efficiency and dispersions of position angles,
we discuss the uniform and random components of the magnetic field along the line-of-sight on the Galactic scale.
Furthermore, we verify whether the wavelength dependence of polarization at NIR is really represented by a power law or not,
and examine what types of dust grains cause the NIR polarization.
In section 2, we describe observations and data reduction.
In section 3, we present results for color excess, polarization, the polarization efficiency,
and the wavelength dependence of polarization in our sample.
In section 4, we discuss the results as they relate to the magnetic field and polarizing grain properties toward the GC.

\section{OBSERVATIONS AND DATA REDUCTION}

Polarimetric imaging observations toward the GC have been carried out with a NIR polarimetric camera
installed on the IRSF (InfraRed Survey Facility) 1.4 m telescope at the South African Astronomical Observatory (SAAO) in Sutherland.
The NIR polarimetric camera consists of the single-beam polarimeter SIRPOL
[a rotating achromatic (1.0$-$2.5 $\micron$) half-wave plate and a wire grid polarizer; \citealp{Kandori06}]
and NIR camera SIRIUS \citep[Simultaneous three-color InfraRed Imager for Unbiased Survey;][]{Nagashima99,Nagayama03}.
The camera is equipped with three 1024 pixel $\times$ 1024 pixel HAWAII arrays.
This enables simultaneous observations in the $J$ (central wavelength $\lambda_J$ = 1.25 $\micron$),
$H$ ($\lambda_H$ = 1.63 $\micron$), and $K_S$ ($\lambda_{K_S}$ = 2.14 $\micron$) bands
by splitting the beam into the three wavelengths with two dichroic mirrors.
The image scale of the arrays is 0$\farcs$45 pixel$^{-1}$, yielding a field of view of 460$\arcsec$ $\times$ 460$\arcsec$.

\begin{figure}
  \epsscale{1.0}
  \plotone{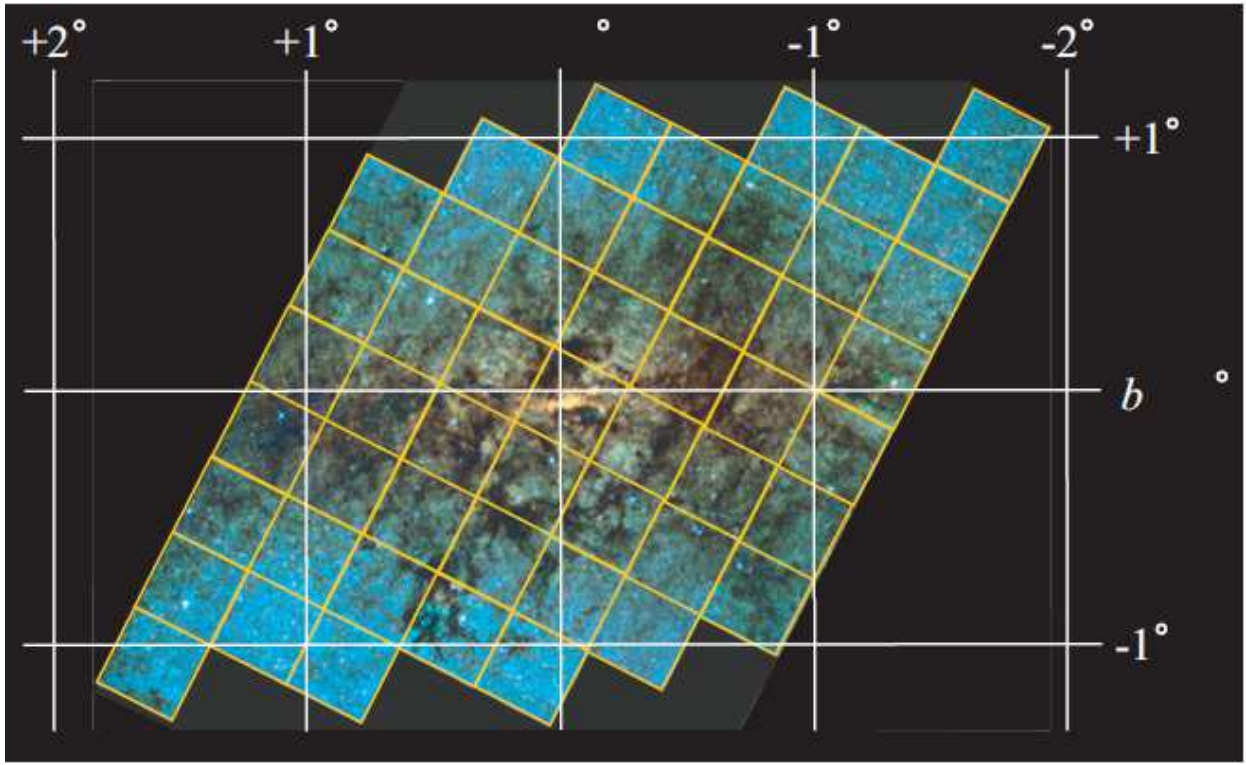}
  \caption{The observed area in this study.
    The background is the $JHK_S$ composite image of the GC \citep[Fig. 1.5 of][]{Nishiyama05}.
    Each square corresponds to 3 $\times$ 3 fields with a size of 20$\arcmin$ $\times$ 20$\arcmin$.}
  \label{area}
\end{figure}

From 2006 to 2009, we have observed 459 fields toward the GC, and the total area covered is about 5.7 deg$^2$
(see Fig. \ref{area}).
The centers of fields were set at intervals of 400$\arcsec$.
We obtained 10 dithered frames on the circle with a radius of 20$\arcsec$, yielding an effective field of view of about 420$\arcsec$
and overlaps between adjacent fields with a size of about 420$\arcsec$ $\times$ 20$\arcsec$.
We performed 10 s exposures at four wave plate angles (0$\arcdeg$, 45$\arcdeg$, 22$\fdg$5, and 67$\fdg$5),
resulting in a total exposure time of 100 s per wave plate angle for each field.
Our observations were carried out under stable sky conditions on photometric nights.
The seeing was typically 1$\farcs$3, 1$\farcs$2, and 1$\farcs$1 (FWHM) in the $J$, $H$, and $K_S$ bands, respectively.
To make median sky frames we observed one of two sparse stellar fields
($l$ = $-$2$\fdg$933, $b$ = 7$\fdg$100; $l$ = 4$\fdg$525, $b$ = $-$12$\fdg$838)
for each one or two field(s), that is, as frequently as every about 10 or 20 minutes.
Twilight flat frames were obtained before and after the observations.
Dark frames were obtained at the end of the nights.
The polarimetric standard star R CrA No. 88 \citep{Whittet92} was observed 15 times through the observing runs,
with 1.6 or 2 s exposures at each wave plate angle at 10 dithered positions.

We applied the standard procedures of NIR array image reduction, including dark-current subtraction, flat-fielding,
sky subtraction, and frame combination using the IRAF (Image Reduction and Analysis Facility)%
\footnote[6]{IRAF is distributed by the National Optical Astronomy Observatories, which are operated by the Association
of Universities for Research in Astronomy, Inc., under cooperative agreement with the National Science Foundation.} software package.
After subtraction of an averaged dark frame, each frame was divided by a normalized flat frame.
Then the thermal emission pattern, the fringe pattern due to OH emission, and the reset-anomaly slope pattern of the HAWAII arrays
were subtracted from each frame with a median sky frame.
This subtraction cannot be done adequately in the case that the intensity of OH emission or temperature abruptly change,
so that we observed the field again.
Finally we obtained images by combining 10 frames at each wave plate angle, and Stokes $I$ images
by combining 10 $\times$ 4 = 40 frames.

Photometry of point sources was performed using the DAOPHOT package in IRAF.
We used the DAOFIND task to detect point sources on Stokes $I$ images.
Since the observed fields are highly crowded stellar fields, 
we obtained positions and magnitudes of detected sources on Stokes $I$ images from PSF-fitting photometry using the ALLSTAR task.
A model PSF was constructed from bright and isolated sources whose numbers were typically 20$-$60, 50$-$90, and 60$-$90
on each image in the $J$, $H$, and $K_S$ bands, respectively.
Positions on celestial coordinate systems were calculated referring to the 2MASS Point Source Catalog%
\footnote[7]{We related pixel coordinates to celestial coordinates using the OPM software, which is compiled by Dr. N. Matsunaga
and based on the Optimistic Pattern Matching algorithm proposed by \citet{Tabur07}.} \citep{Skrutskie06}.
The positional accuracy was estimated to be about 0$\farcs$03.
For photometric calibration, we compared magnitudes for the detected sources with
those for the 2MASS Point Source Catalog sources in the observed field.
In this comparison magnitude transformations from the 2MASS system to the IRSF system were applied
(Y. Nakajima 2007, private communication).
The photometric accuracy is about 0.03 mag in all the $JHK_S$ bands.

In order to estimate the effect of confusion on the PSF-fitting photometry in crowded fields,
we computed the completeness of the PSF-fitting photometry.
We added 900 artificial sources with the brightness corresponding to the limiting magnitudes, $J$ $\sim$ 14.0 mag, $H$ $\sim$ 13.4 mag,
and $K_S$ $\sim$ 12.5 mag (see the last paragraph in this section), on each Stokes $I$ image
in a reticular pattern at intervals of 14$\arcsec$ (30 $\times$ 30 sources) using the ADDSTAR task,
and performed source detection and PSF-fitting photometry in the same manner as above.
The recovery rates of the added artificial sources and median differences of magnitudes
between the recovered and added artificial sources were calculated in each field.
The recovery rates were 98.5\%, 95.9\%, and 95.6\% on average, and the means of the absolute median differences of magnitudes were
0.005 mag, 0.011 mag, and 0.009 mag in the $J$, $H$, and $K_S$ bands, respectively.

In order to calculate the Stokes parameters for sources, we performed detection and aperture photometry of sources
at each wave plate angle using the DAOFIND and PHOT tasks, and obtained flux and positions.
We compared aperture photometry with PSF-fitting photometry by calculating normalized Stokes parameters $Q/I$ and $U/I$
for duplicate sources in overlapping regions of adjacent fields observed under different observing conditions.
As a result, aperture photometry gives a better result than PSF-fitting photometry which shows systematic offsets.
The center of aperture was determined as the centroid of each source independently on images at each wave plate angle.
We adopted the aperture size of 1.5 $\times$ FWHM measured on Stokes $I$ images.
The aperture size was chosen after search of 1.0, 1.5, and 2.0 $\times$ FWHM.
The normalized Stokes parameters were best matched for a factor of 1.5 with each other between overlapping regions
(see below for the comparison of normalized Stokes parameters in overlapping regions).

We also computed the completeness of the aperture photometry in a similar manner to the PSF-fitting photometry.
We added 900 artificial sources with the brightness corresponding to the limiting magnitudes on each image at each wave plate angle
in a reticular pattern at intervals of 14$\arcsec$ (30 $\times$ 30 sources),
and performed source detection and aperture photometry as described above.
We calculated the recovery rates of the added artificial sources and standard deviations of median flux of
the recovered artificial sources among all wave plate angles in each field.
The average recovery rates were 97.5\%, 94.4\%, and 94.3\%, and the standard deviations of the median flux were
0.2\%, 0.3\%, and 0.2\% on average in the $J$, $H$, and $K_S$ bands, respectively.

We merged sources on Stokes $I$ image from PSF-fitting photometry and sources on images at each wave plate angle from aperture photometry.
Positions of sources obtained with PSF-fitting photometry on Stokes $I$ images were used as the reference positions for merging.
We matched sources on Stokes $I$ image and images at each wave plate angle to the reference positions within a 1 pixel radius.
In crowded fields, PSF-fitting photometry yields more precise positions of sources than DAOFIND.
We mitigated source confusion by using the precise reference positions for matching.
Based on flux of sources at each wave plate angle, we calculated Stokes parameters
$I$, $Q$, $U$, and their statistical errors (calculated from noise of signal, sky background, dark current, and readout).
The Stokes parameter $I$ was calculated from the flux on images at each wave plate angle,
as well as the Stokes parameters $Q$ and $U$, not from the flux from the PSF-fitting photometry on Stokes $I$ images,
in order to minimize error due to variation of PSF among images at each wave plate angle and Stokes $I$ images.
These Stokes parameters also have systematic errors originating from change of atmospheric conditions
(seeing and transparency of atmosphere) at each wave plate angle for each field.
To estimate these systematic errors we have made a comparison of normalized Stokes parameters $Q/I$ and $U/I$ of the same sources
in overlapping regions between adjacent fields with sizes of about 420$\arcsec$ $\times$ 20$\arcsec$.
We defined the same sources as the sources having closest positions within 1$\arcsec$.
Using the sources whose statistical errors of $Q/I$ and $U/I$ were less than 1\%,
average differences of $Q/I$ and $U/I$ were calculated in each overlapping region.
The numbers of the same sources in each overlapping region were typically 10$-$50, 40$-$100, and 40$-$100 in the $J$, $H$,
and $K_S$ bands, respectively.
The means of the average differences of $Q/I$ and $U/I$ in overlapping regions were 0.33\% and 0.35\% in the $J$ band,
0.26\% and 0.27\% in the $H$ band, and 0.25\% and 0.26\% in the $K_S$ band.
We adopted these values for the systematic errors of the normalized Stokes parameters $Q/I$ and $U/I$,
and computed the total errors by combining the estimated systematic errors with the statistical errors in quadrature.
Then the observed degree of polarization $p_\mathrm{obs}$, position angle $\theta$
and their errors $\delta p$ and $\delta \theta$ were derived.
$\theta$ is defined as the angle between \textbf{\textit{E}}-vector of polarization and the direction of north celestial pole
and increasing to the east ($-$90$\arcdeg$ $\leq$ $\theta$ $<$ 90$\arcdeg$).
To correct noise biasing we calculated degree of polarization $p$ using the following formula:
\begin{displaymath}
  p = \sqrt{{p_\mathrm{obs}}^2 - \delta p^2}
\end{displaymath}
\citep{Wardle74,Clarke86}.
We regarded sources with $p_\mathrm{obs}$ $\leq$ $\delta p$ as unpolarized ($p$ = 0) sources.
All the data were calibrated for the polarization efficiency of the wave plate and polarizer
\citep[95.5\%, 96.3\%, and 98.5\% in the $J$, $H$, and $K_S$ bands; see][]{Kandori06}.

We checked our analysis by comparing our polarimetry of the polarimetric standard star R CrA No. 88 with that by \citet{Whittet92}.
The values $p$ and $\theta$ were derived by averaging normalized Stokes parameters $Q/I$ and $U/I$
obtained from 15 observations.
Their errors $\delta p$ and $\delta \theta$ were determined from the standard deviations of the means of $Q/I$ and $U/I$.
As shown in Table \ref{standard}, we confirmed that our polarimetry is consistent with that of \citet{Whittet92}
within the errors in all the bands.

\begin{figure}
  \epsscale{1.0}
  \plotone{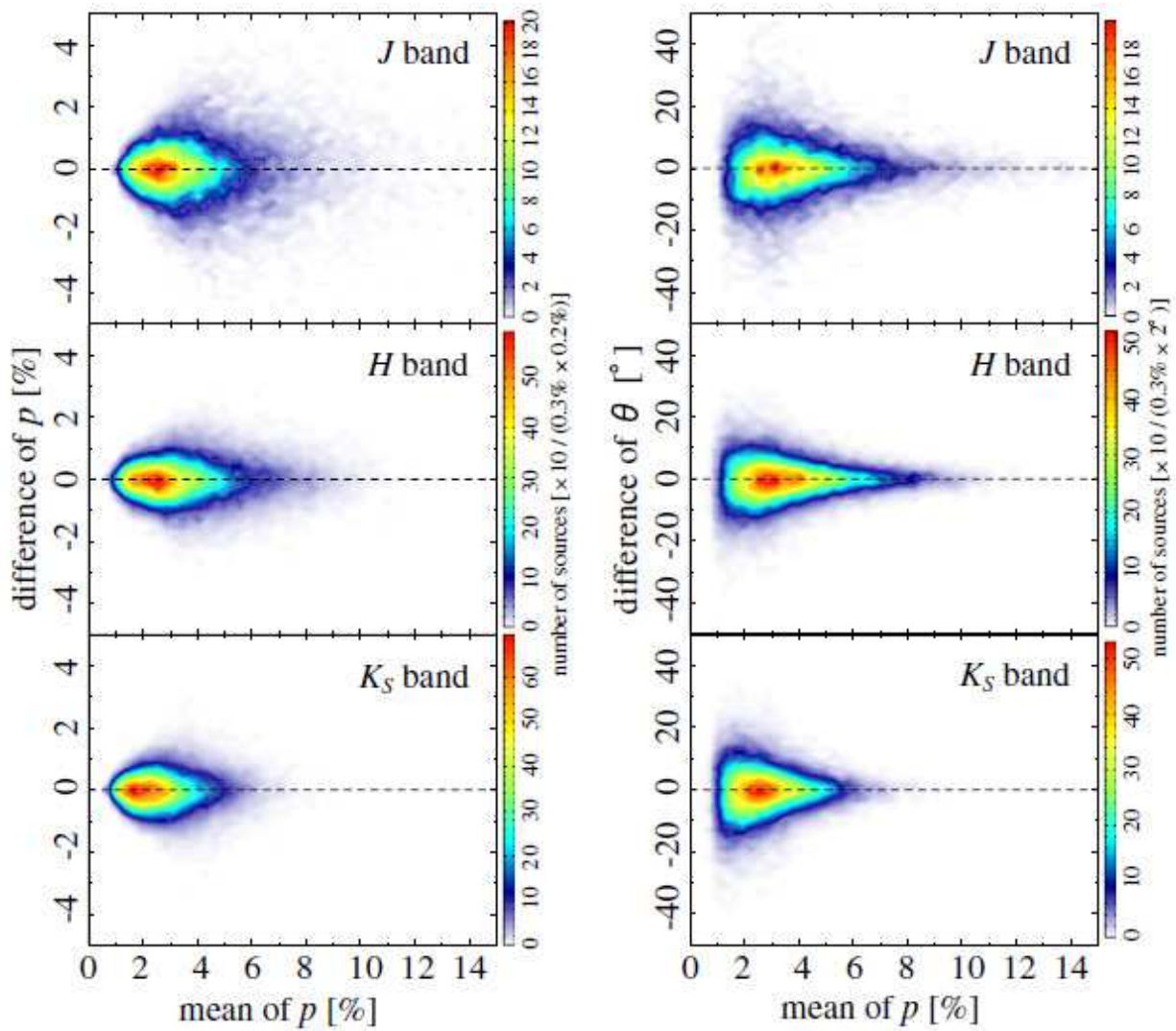}
  \caption{Differences of $p$ ($left$ $side$) and $\theta$ ($right$ $side$) as a function of means of $p$ in the $J$ ($top$),
    $H$ ($middle$), and $K_S$ ($bottom$) bands based on the comparison of the same sources with $\delta p$ $\leq$ 1\%
    and $\delta \theta$ $\leq$ 10$\arcdeg$ in overlapping regions.}
  \label{pcmp2}
\end{figure}

For an additional check of our analysis, we made a comparison of $p$ and $\theta$ of the same sources in overlapping regions
in a similar way to the comparison of $Q/I$ and $U/I$.
Using the same sources with $\delta p$ $\leq$ 1\% and $\delta \theta$ $\leq$ 10$\arcdeg$,
we calculated differences of $p$ and $\theta$.
The numbers of the same sources in each overlapping region were typically 10$-$30, 20$-$80, and 10$-$70 in the $J$, $H$,
and $K_S$ bands, respectively.
Figure \ref{pcmp2} shows differences of $p$ and $\theta$ as a function of means of $p$ for the sources.
In all the range of the measured polarization, the absolute differences of $p$ are mostly less than 1\%,
and those of $\theta$ are mostly less than 10$\arcdeg$;
the standard deviations of the differences of $p$ are 0.79\%, 0.55\%, and 0.49\%,
and those of $\theta$ are 6$\fdg$9, 5$\fdg$3, and 6$\fdg$3 in the $J$, $H$, and $K_S$ bands, respectively

Source detections at $J$, $H$, and $K_S$ were merged into a single source record using the positions in each band.
First the $K_S$ sources were taken as seed detections; then the $K_S$-$H$ pairwise matching was done,
and this was followed by the $K_S$-$J$ pairwise matching.
Then the $H$ sources which were not matched in the previous process were taken as seeds;
then the $H$-$J$ pairwise matching was done.
The match was acceptable if the source separation was less than 1$\arcsec$.
We adopted the coordinates for the longest wavelength in the matching as the source coordinates.
There are a total of 3,539,087 sources:
1,536,017, 2,979,994, and 3,190,511 sources detected in the $J$, $H$, and $K_S$ bands, respectively.
The limiting magnitudes of our survey, defined as the level at which $\delta p$ $\leq$ 1\%,
are $J$ $\sim$ 14.0 mag, $H$ $\sim$ 13.4 mag, and $K_S$ $\sim$ 12.5 mag.
There are 234,121 sources with $\delta p_J$ $\leq$ 1\%, 541,990 sources with $\delta p_H$ $\leq$ 1\%,
and 558,647 sources with $\delta p_{K_S}$ $\leq$ 1\%.

\section{RESULTS}

\subsection{MK CLASSIFICATIONS AND LOCATIONS}
\label{MK CLASSIFICATIONS AND LOCATIONS}

\begin{figure}
  \epsscale{0.9}
  \plotone{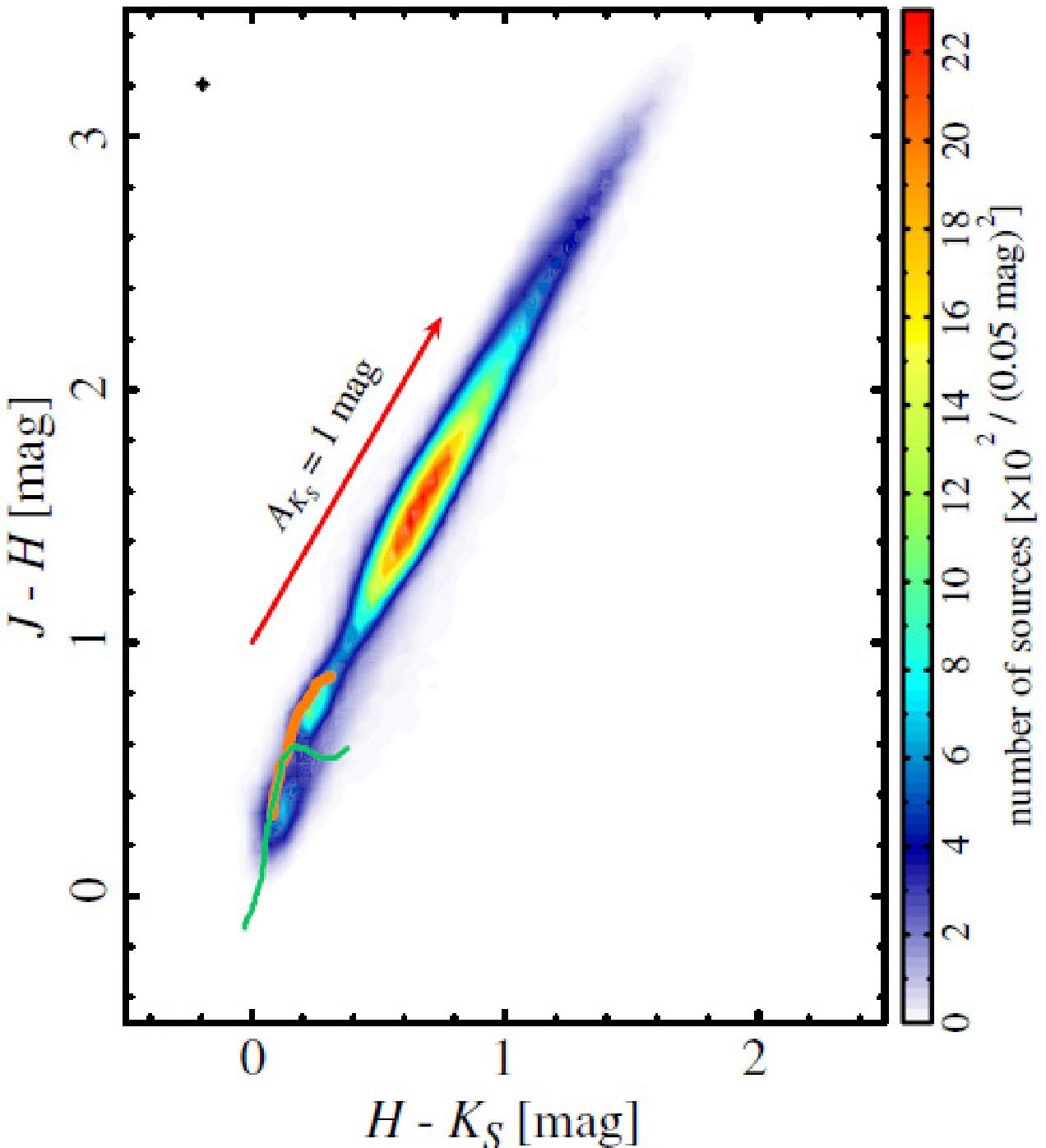}
  \caption{$J - H$ vs. $H - K_S$ color-color diagram for the sources that are detected in all the bands
    and have $\delta p$ $\leq$ 1\% in all the bands.
    The thin and thick curves are the loci of dwarfs and giants, respectively.
    The data for O9$-$B9 dwarfs are from \citet{Koornneef83}, and those for A0$-$M6 dwarfs and G0$-$M7 giants are from
    \citet{Bessell88}.
    The arrow indicates a reddening vector whose slope is 1.72 \citep{Nishiyama06},
    and its length corresponds to extinction of $A_{K_S}$ = 1 mag.
    The upper left cross denotes the average errors of colors for the sources.}
  \label{ccd}
\end{figure}

A color-color diagram for the sources is shown in Figure \ref{ccd}.
The figure includes 165,858 sources that have $\delta p$ $\leq$ 1\% in all the bands.
There are three distinct populations; the majority of the sources fall in a feature
extending parallel to the reddening vector from the locus of giants, while two weak concentrations of the sources are seen
around ($H - K_S$, $J - H$)$\sim$(0.1, 0.3) mag and $\sim$(0.2, 0.7) mag along the loci of dwarfs and giants.
Based on the model by \citet{Wainscoat92}, we estimate what kind of sources can be detected in all the bands
within the limiting magnitudes ($J$ $\sim$ 14.0 mag, $H$ $\sim$ 13.4 mag, and $K_S$ $\sim$ 12.5 mag).
The model predicts the numbers of sources expected for each MK classification (spectral type and luminosity class)
at each distance from the Sun.
Toward the GC, we would expect that most of the detectable sources are K/M giants located in the Galactic bulge.
They are heavily reddened due to the large amount of the intervening dust, corresponding to the extended feature in Figure \ref{ccd}.
Among the sources located in the Galactic disk ($\lesssim$ 4 kpc from the Sun), A/F dwarfs and G/K giants are mainly detected.
The two concentrations around the loci of dwarfs and giants predominantly consist of these dwarfs and giants with small extinction.

Out of 234,121 sources with $\delta p_J$ $\leq$ 1\%, 541,990 sources with $\delta p_H$ $\leq$ 1\%,
and 558,647 sources with $\delta p_{K_S}$ $\leq$ 1\%,
those having $H - K_S$ colors amount to 196,651, 512,030, and 544,675 sources in the $J$, $H$, and $K_S$ bands, respectively.
We divide these sources into two groups at $H - K_S$ $\sim$ 0.4 mag,
where a saddle of the extended feature of giants exists;
the disk sources have $H - K_S$ $<$ 0.4 mag and bulge sources have $H - K_S$ $\geq$ 0.4 mag.
The former are mostly A/F dwarfs and G/K giant located in the Galactic disk,
while the latter are mostly K/M giants located in the Galactic bulge.
The numbers of the disk and bulge sources amount to 58,007 and 138,644 with $\delta p_J$ $\leq$ 1\%,
54,102 and 457,928 with $\delta p_H$ $\leq$ 1\%, and 31,448 and 513,227 with $\delta p_{K_S}$ $\leq$ 1\%.
Hereafter we call these sources as the disk and bulge sources.

\subsection{COLOR EXCESS}

We calculate color excess for the disk and bulge sources using the equation:
\begin{displaymath}
  E(H - K_S) = (H - K_S) - \langle(H - K_S)_0\rangle.
\end{displaymath}
The mean of the intrinsic colors of sources, $\langle$$(H - K_S)_0$$\rangle$, is computed as follows.
First we compute the numbers of sources expected in each band for each MK classification in the disk and bulge
based on the model by \citet{Wainscoat92} under the following criteria.
The criteria for the disk and bulge sources are detection within the limiting magnitudes in each band, that is,
detection with $J$ $\lesssim$ 14.0 mag, $H$ $\lesssim$ 13.4 mag, and $K_S$ $\lesssim$ 12.5 mag.
Additional criterion for the disk sources is $H - K_S$ $<$ 0.4 mag, and that for the bulge sources is $H - K_S$ $\geq$ 0.4 mag.
We compute $\langle$$(H - K_S)_0$$\rangle$ by averaging the intrinsic colors of sources, $(H - K_S)_\mathrm{C0}$, using the equation:
\begin{displaymath}
  \langle(H - K_S)_0\rangle = \frac{\displaystyle \sum_\mathrm{C} \{(H - K_S)_\mathrm{C0} \times \mathrm{N_C}\}}{\displaystyle \sum_\mathrm{C} \mathrm{N_C}},
\end{displaymath}
where the sum is over the 29 spectral classes in the \citet{Wainscoat92} model with luminosity classes of I\hspace{-.1em}I\hspace{-.1em}I and V
(see their Table 2), N$_\mathrm{C}$ is the total number of sources of class C predicted by the model, and the intrinsic colors,
$(H - K_S)_\mathrm{C0}$, for each class are taken from \citet{Koornneef83} and \citet{Bessell88}.
For the uncertainty of $\langle$$(H - K_S)_0$$\rangle$, we take the standard deviation of the intrinsic colors of sources from an equation:
\begin{displaymath}
  \sigma((H - K_S)_0) = \sqrt{\frac{\displaystyle \sum_\mathrm{C} [\{(H - K_S)_\mathrm{C0} - \langle(H - K_S)_0\rangle\}^2 \times \mathrm{N_C}]}{\displaystyle \sum_\mathrm{C} \mathrm{N_C}}}.
\end{displaymath}
$\langle$$(H - K_S)_0$$\rangle$ and $\sigma((H - K_S)_0)$ in an area with a size of 10$\arcmin$ $\times$ 10$\arcmin$ are computed
in each direction of the disk and bulge sources.
From line-of-sight to line-of-sight, $\langle$$(H - K_S)_0$$\rangle$ and $\sigma((H - K_S)_0)$ vary,
because of the spatial variations of number density of sources with each MK classification in the Wainscoat's model.
As for the disk sources, $\langle$$(H - K_S)_0$$\rangle$ and $\sigma((H - K_S)_0)$ are computed to be 0.06$-$0.08 mag and 0.05 mag
in the $J$ band, 0.07$-$0.08 mag and 0.04$-$0.05 mag in the $H$ band, and 0.08$-$0.10 mag and 0.04$-$0.05 mag in the $K_S$ band
for lines of sight toward each disk source.
As for the bulge sources, $\langle$$(H - K_S)_0$$\rangle$ and $\sigma((H - K_S)_0)$ are computed to be 0.13$-$0.19 mag and 0.05 mag
in the $J$ band, 0.16$-$0.21 mag and 0.04$-$0.08 mag in the $H$ band, and 0.17$-$0.22 mag and 0.04$-$0.06 mag in the $K_S$ band
for lines of sight toward each bulge source.

\begin{figure}
  \epsscale{0.4}
  \plotone{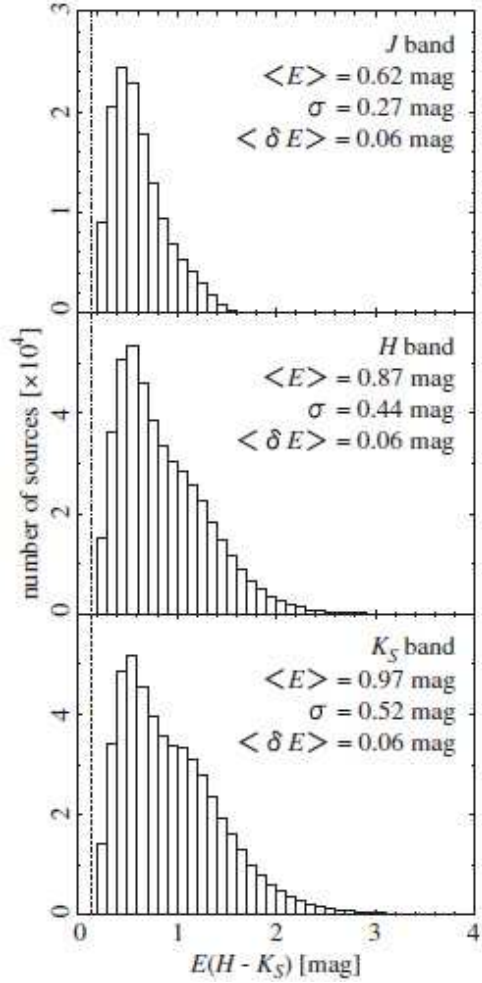}
  \caption{Histograms of color excess $E(H - K_S)$ for the bulge sources with $\delta p_J$ $\leq$ 1\% ($top$),
    those with $\delta p_H$ $\leq$ 1\% ($middle$), and those with $\delta p_{K_S}$ $\leq$ 1\% ($bottom$).
    The means, standard deviations, and average errors of $E(H - K_S)$ for the bulge sources are shown
    at the upper right of the panels.
    The dash-dotted lines in each panel show the means of $E(H - K_S)$ for the disk sources with $\delta p$ $\leq$ 1\% in each band.}
  \label{exthistogram}
\end{figure}

The color excess $E(H - K_S)$ for a disk source is not so large compared with its error,
and therefore cannot be accurately determined.
We adopt only means of $E(H - K_S)$ for the disk sources for further discussions.
From the means of $H - K_S$ (0.20 mag, 0.21 mag, and 0.22 mag) and average intrinsic color of the disk sources
(0.07 mag, 0.08 mag, and 0.09 mag),
the means of $E(H - K_S)$ are calculated to be 0.13 mag, 0.13 mag, and 0.13 mag for the disk sources
in the $J$, $H$, and $K_S$ bands, respectively (dash-dotted lines in Fig. \ref{exthistogram}).

Contrary to $E(H - K_S)$ for a disk source, $E(H - K_S)$ for a bulge source is so large that
it can be accurately determined.
We show histograms of $E(H - K_S)$ for the bulge sources in Figure \ref{exthistogram}.
The distributions of $E(H - K_S)$ peak at about 0.5 mag and have tails toward large color excess in all the bands.
The bulge sources in the $H$ and $K_S$ bands trace larger color excess than those in the $J$ band
because of the wavelength dependence of interstellar extinction.

\begin{figure}
  \epsscale{1.0}
  \plotone{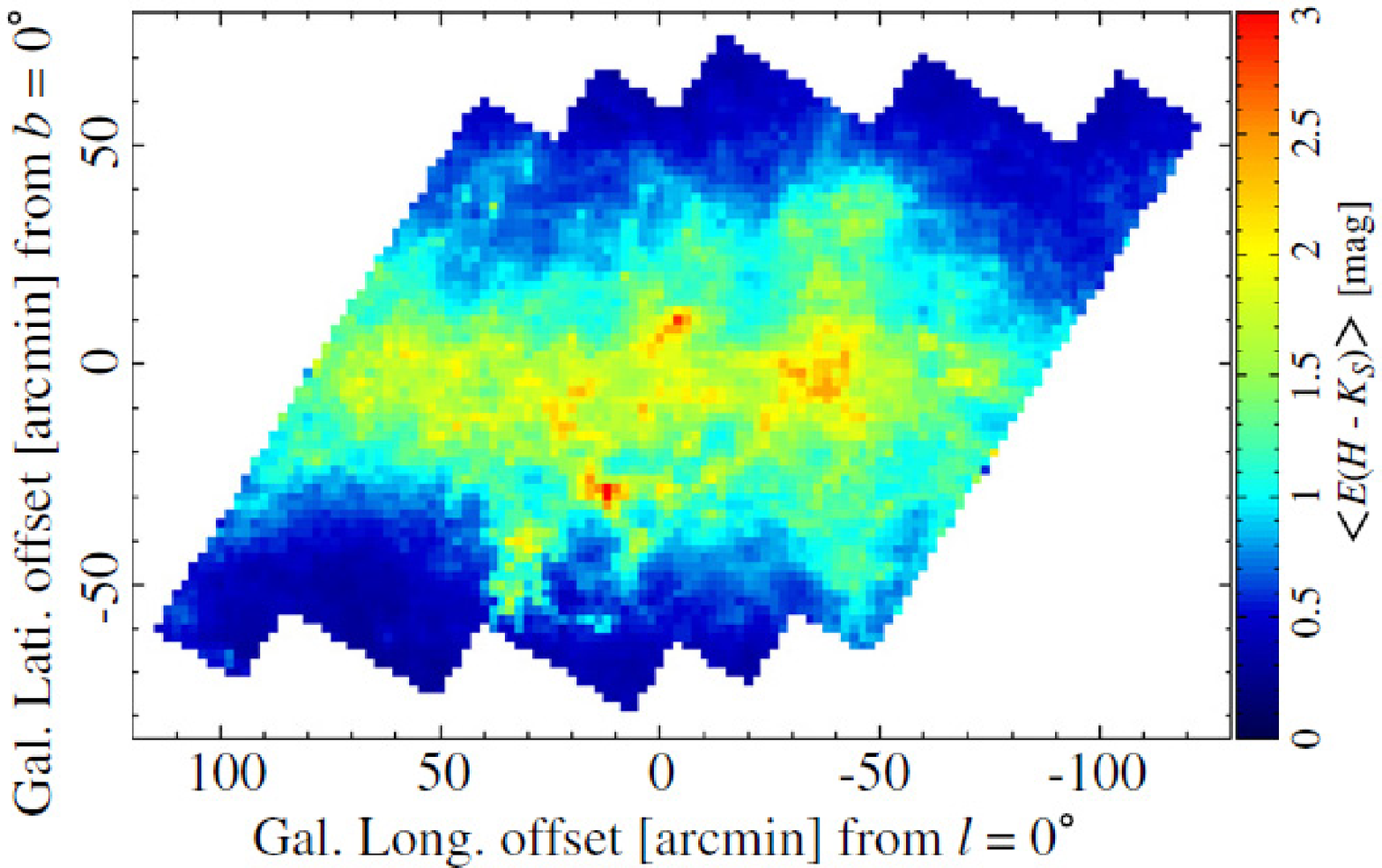}
  \caption{Map of $E(H - K_S)$.
    Each pixel represents a mean of $E(H - K_S)$ for the bulge sources with $\delta p_{K_S}$ $\leq$ 1\% in each cell
    with a size of 2$\arcmin$ $\times$ 2$\arcmin$.}
  \label{kexttile}
\end{figure}

We present a map of $E(H - K_S)$ for the bulge sources in the $K_S$ band as follows.
The observed area is divided into cells with a size of 2$\arcmin$ $\times$ 2$\arcmin$.
Means of $E(H - K_S)$ for the bulge sources in the $K_S$ band are calculated in each cell.
The size of cells is determined to contain as many sources in each cell as possible without degradation of the angular resolution.
The resultant map of the mean $\langle$$E(H - K_S)$$\rangle$ is shown in Figure \ref{kexttile}.
The map shows the spatial variation of interstellar extinction depending on the Galactic latitude.
The cells with large values $\langle$$E(H - K_S)$$\rangle$ $\geq$ 2 mag mainly concentrate to the Galactic plane
($|b|$ $\leq$ 0$\fdg$5), showing clumpy and filamentary structures.

\subsection{DEGREE OF POLARIZATION AND POSITION ANGLE}
\label{DEGREE OF POLARIZATION AND POSITION ANGLE}

\begin{figure}
  \epsscale{1.0}
  \plotone{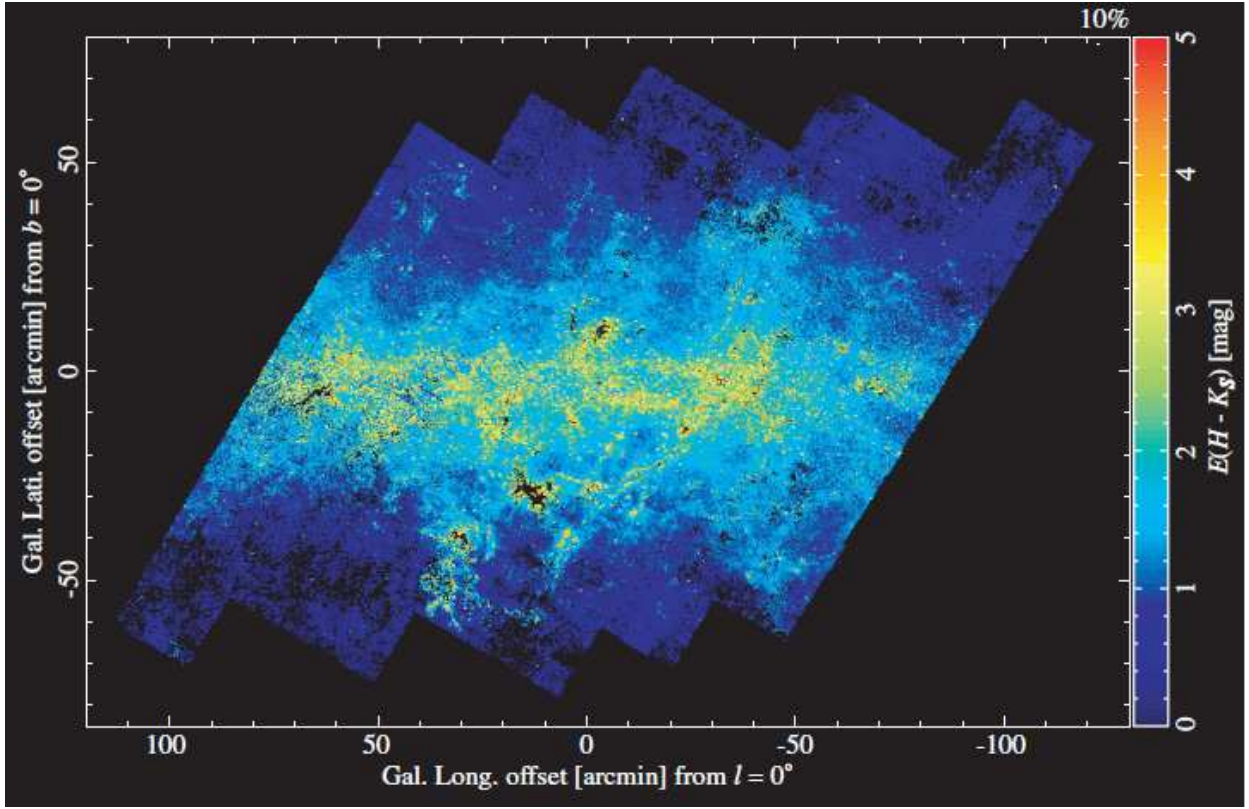}
  \caption{$K_S$ band polarization vector map for the disk and bulge sources with $\delta p_{K_S}$ $\leq$ 1\%
    and $\delta \theta_{K_S}$ $\leq$ 10$\arcdeg$.
    Each bar is parallel to the \textbf{\textit{E}}-vector of polarization.
    The length of each bar is proportional to degree of polarization.
    Color excess for the bulge sources is also shown as color of the bars.
    The disk sources are shown by white bars.
    Each bar including the bar for 10\% scale of polarization degree at the upper right can be recognized by zooming the figure
    in the electronic edition.}
  \label{kpextvector}
\end{figure}

We plot a $K_S$ band polarization vector map for the disk and bulge sources in Figure \ref{kpextvector}.
Zooming the electronic edition of the map makes the \textbf{\textit{E}}-vectors of polarization for sources legible.
Most of the sources show the \textbf{\textit{E}}-vectors of polarization nearly parallel to the Galactic plane,
while there exist some deviations.
These deviations of \textbf{\textit{E}}-vectors are seen in the regions
where a relatively few sources with large color excess are detected.
Most of the detected sources with large color excess could be in/behind the nearby dense clouds.
The deviated \textbf{\textit{E}}-vectors would reflect the local magnetic field directions in nearby dense clouds.

\begin{figure}
  \epsscale{0.5}
  \plotone{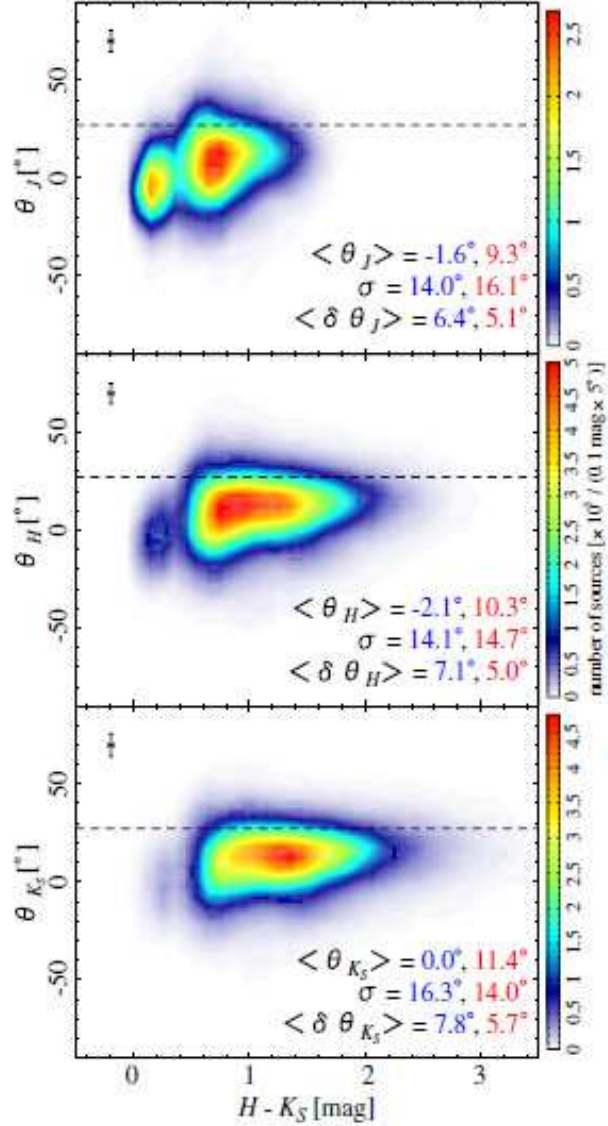}
  \caption{Position angles $\theta$ vs. $H - K_S$ colors for the disk and bulge sources with $\delta p_J$ $\leq$ 1\%
    and $\delta \theta_J$ $\leq$ 10$\arcdeg$ ($top$), those with $\delta p_H$ $\leq$ 1\% and $\delta \theta_H$ $\leq$ 10$\arcdeg$
    ($middle$), and those with $\delta p_{K_S}$ $\leq$ 1\% and $\delta \theta_{K_S}$ $\leq$ 10$\arcdeg$ ($bottom$).
    The dashed lines in each panel represent the orientation of Galactic plane ($\sim$27$\arcdeg$).
    The means, standard deviations, and average errors of $\theta$ for the disk ($left$ $values$)
    and bulge ($right$ $values$) sources are shown at the lower right of the panels.
    The upper left crosses in each panel denote the average errors of $\theta$ and $H - K_S$ for the sources.}
  \label{pacolor}
\end{figure}

Figure \ref{pacolor} shows the relations between position angles and colors for the disk and bulge sources
with $\delta \theta$ $\leq$ 10$\arcdeg$ in each band.
There are two distinct populations in blue ($H - K_S$ $<$ 0.4 mag) and red ($H - K_S$ $\geq$ 0.4 mag) colors,
corresponding to the disk and bulge sources, respectively (see \S\ref{MK CLASSIFICATIONS AND LOCATIONS} and Fig. \ref{ccd}).
The disk sources are prominent in the $J$ band, but less prominent in the $H$ and $K_S$ bands.
The numbers of the disk and bulge sources amount to 38,425 and 118,045 with $\delta \theta_J$ $\leq$ 10$\arcdeg$,
20,858 and 355,124 with $\delta \theta_H$ $\leq$ 10$\arcdeg$, and 6,086 and 323,603 with $\delta \theta_{K_S}$ $\leq$ 10$\arcdeg$.
The means of $\theta$ clearly differ between the two populations as also noted by \citet{Kobayashi83} and \citet{Nishiyama09a};
the means of $\theta$ are 9$\fdg$3, 10$\fdg$3, and 11$\fdg$4 for the bulge sources,
while $-$1$\fdg$6, $-$2$\fdg$1, and 0$\fdg$0 for the disk sources in the $J$, $H$, and $K_S$ bands, respectively.
Most of the bulge sources have $\theta$ nearly parallel to the Galactic plane ($\sim$27$\arcdeg$), but slightly rotated westward.
Position angles rotate more for the disk sources.

\begin{figure}
  \epsscale{1.0}
  \plotone{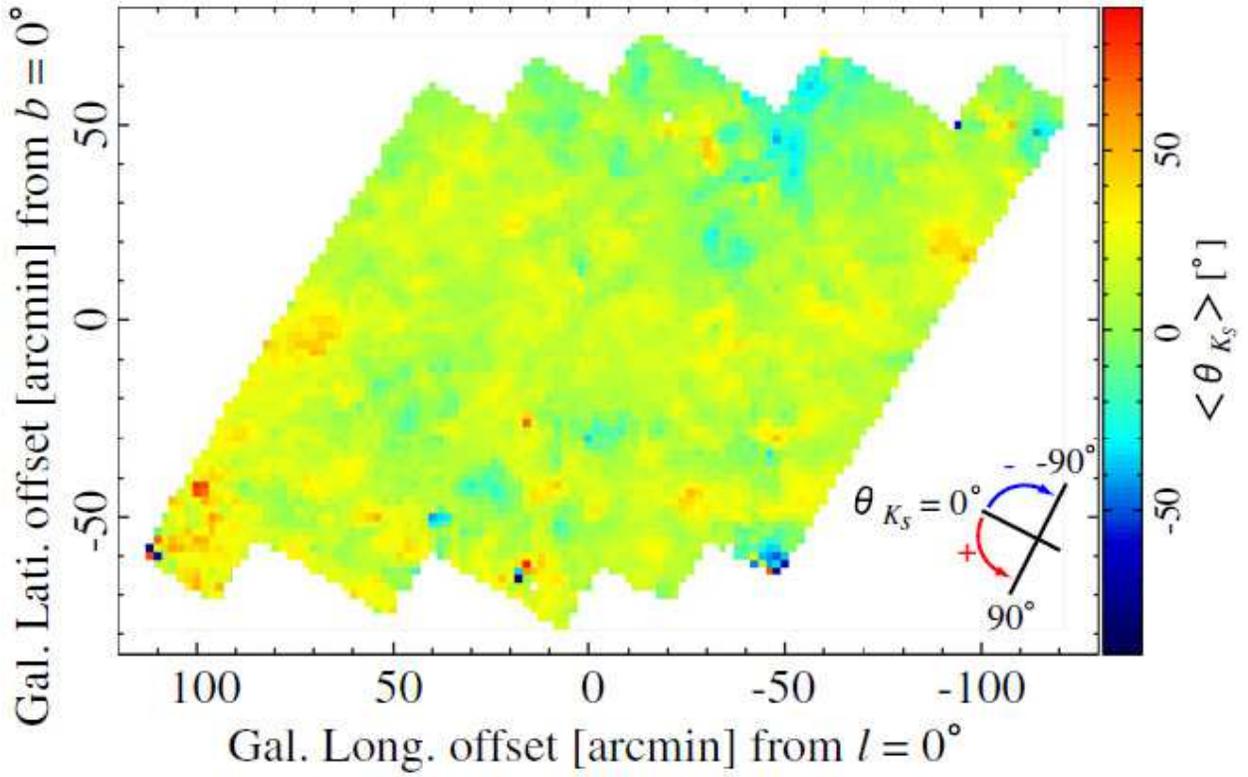}
  \caption{Map of $\theta_{K_S}$.
    Each pixel represents a mean of $\theta_{K_S}$ for the bulge sources with $\delta p_{K_S}$ $\leq$ 1\%
    and $\delta \theta_{K_S}$ $\leq$ 10$\arcdeg$ in each cell with a size of 2$\arcmin$ $\times$ 2$\arcmin$.
    The white pixels include no sources and means cannot be measured.}
  \label{kpatile}
\end{figure}

In a similar manner as $E(H - K_S)$, we present a map of $\theta_{K_S}$ for the bulge sources
with $\delta \theta_{K_S}$ $\leq$ 10$\arcdeg$ in Figure \ref{kpatile}.
Most of cells show the values 0$\arcdeg$ $\lesssim$ $\langle$$\theta_{K_S}$$\rangle$ $\lesssim$ 20$\arcdeg$,
which indicate that the magnetic fields between the GC and us are longitudinal on average.
These values are close to the average position angle which traces the magnetic field configuration in the GC
\citep[16.0$\arcdeg$;][]{Nishiyama09a}, suggesting that the Galactic magnetic field running nearly parallel to the spiral arms
would connect to the toroidal magnetic field in the GC \citep[see also][]{Novak03,Chuss03}.

\begin{figure}
  \epsscale{0.5}
  \plotone{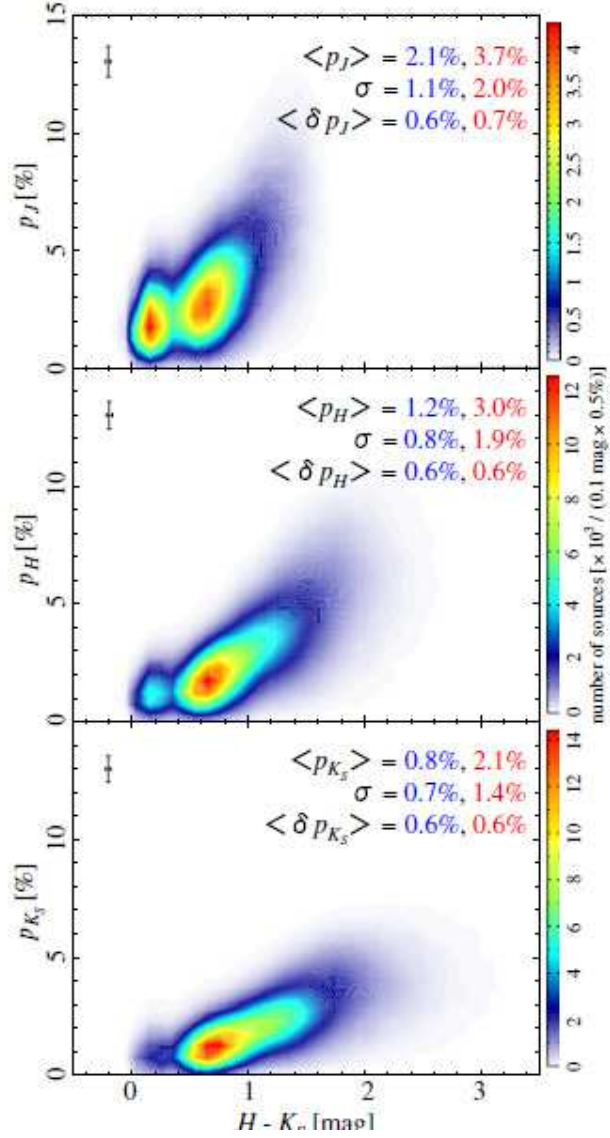}
  \caption{Degrees of polarization $p$ vs. $H - K_S$ colors for the disk and bulge sources
    with $\delta p_J$ $\leq$ 1\% ($top$), those with $\delta p_H$ $\leq$ 1\% ($middle$),
    and those with $\delta p_{K_S}$ $\leq$ 1\% ($bottom$).
    The means, standard deviations, and average errors of $p$ for the disk ($left$ $values$)
    and bulge ($right$ $values$) sources are shown at the upper right of the panels.
    The upper left crosses in each panel denote the average errors of $p$ and $H - K_S$ for the sources.}
  \label{pcolor}
\end{figure}

In the relations between polarization degrees and colors for the disk and bulge sources (Fig. \ref{pcolor}),
two distinct populations corresponding to the disk and bulge sources can be also seen.
The disk sources have average polarization degrees of 2.1\%, 1.2\%, and 0.8\% in the $J$, $H$, and $K_S$ bands, respectively.
Meanwhile, the bulge sources extend red-ward
and show correlations between $p$ and $H - K_S$ in all the bands; $p$ increases with increasing $H - K_S$.
The slopes of the correlations correspond to the polarization efficiency (\S\ref{THE POLARIZATION EFFICIENCY}),
and the difference of the slopes between the bands is due to the wavelength dependence of interstellar polarization
(\S\ref{THE WAVELENGTH DEPENDENCE OF POLARIZATION}).

\begin{figure}
  \epsscale{1.0}
  \plotone{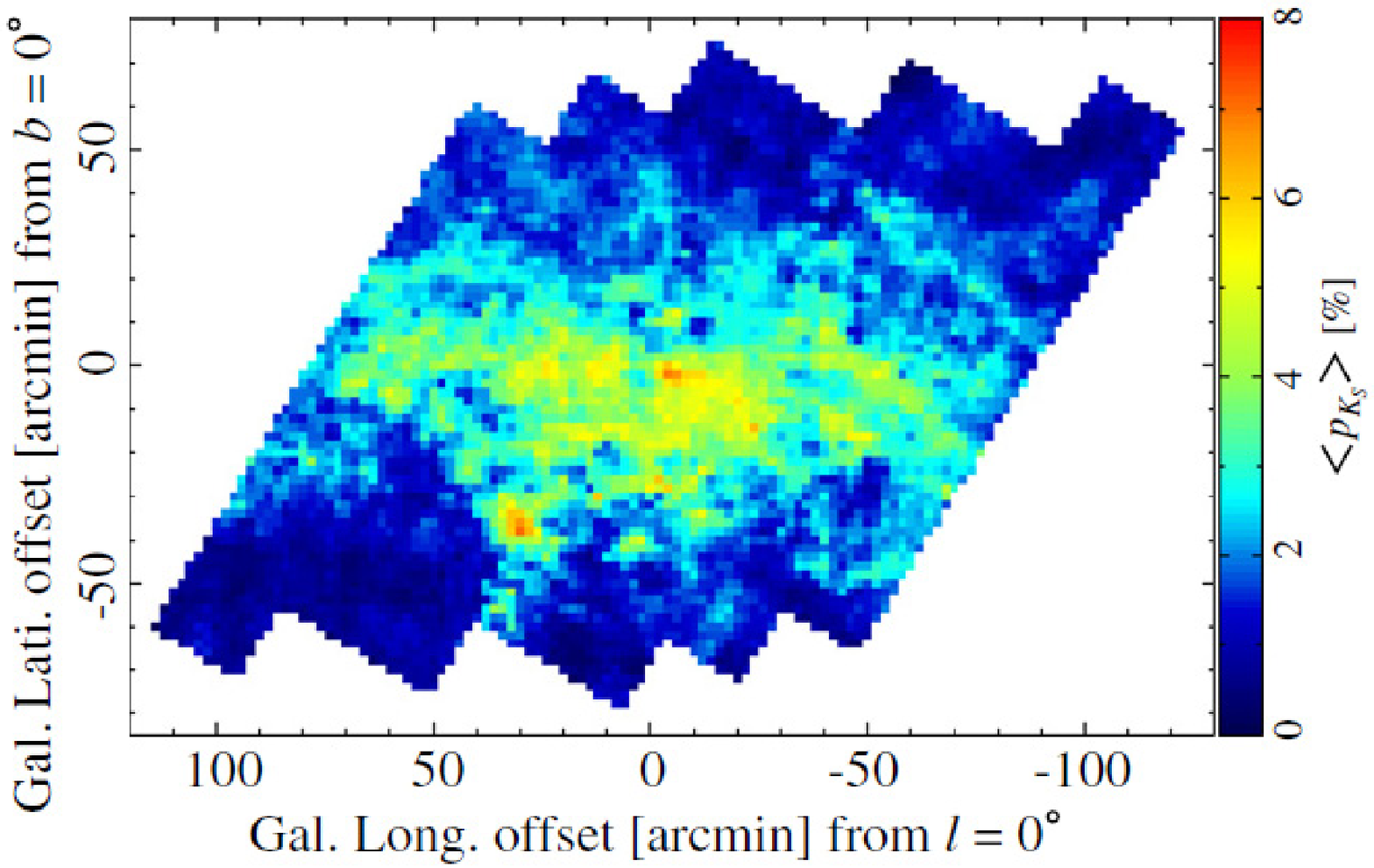}
  \caption{Map of $p_{K_S}$.
    Each pixel represents a mean of $p_{K_S}$ for the bulge sources with $\delta p_{K_S}$ $\leq$ 1\%
    in each cell with a size of 2$\arcmin$ $\times$ 2$\arcmin$.}
  \label{kptile}
\end{figure}

In Figure \ref{kptile}, we show a map of the mean $\langle$$p_{K_S}$$\rangle$.
In a similar way to the spatial variation of $E(H - K_S)$ (Fig. \ref{kexttile}), that of $p_{K_S}$ is dependent on
the Galactic latitude.
Most of cells in which $\langle$$p_{K_S}$$\rangle$ exceeds 5\% are close to the Galactic plane,
while in cells at higher Galactic latitude $\langle$$p_{K_S}$$\rangle$ is only 1$-$2\% or less.
However, the spatial variations of $E(H - K_S)$ and $p$ do not completely coincide with each other.
This is more obvious in the form of the spatial variation of the polarization efficiency (\S\ref{THE POLARIZATION EFFICIENCY}).

\subsection{THE POLARIZATION EFFICIENCY}
\label{THE POLARIZATION EFFICIENCY}

Starlight suffers both extinction and polarization in the passage through the intervening ISM.
Starlight from the disk sources passes through the ISM in the disk ($\lesssim$ 4 kpc from the Sun),
whereas starlight from the bulge sources passes through the ISM between the GC and us
(i.e, the ISM in the disk and bulge; see \S\ref{MK CLASSIFICATIONS AND LOCATIONS}).
From $p/E(H - K_S)$ for the disk and bulge sources,
we examine the polarization efficiency of the ISM in the disk and that between the GC and us.

\begin{figure}
  \epsscale{0.4}
  \plotone{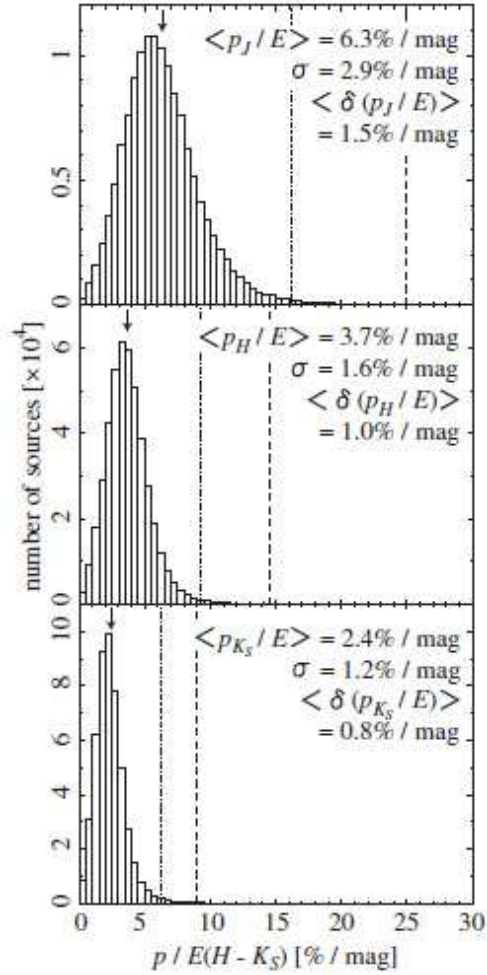}
  \caption{Histograms of the polarization efficiency $p/E(H - K_S)$ for the bulge sources with $\delta p_J$ $\leq$ 1\% ($top$),
    those with $\delta p_H$ $\leq$ 1\% ($middle$), and those with $\delta p_{K_S}$ $\leq$ 1\% ($bottom$).
    The dashed lines in each panel are $p_J/E(H - K_S)$ = 25.0\% / mag, $p_H/E(H - K_S)$ = 14.5\% / mag,
    and $p_{K_S}/E(H - K_S)$ = 9.0\% / mag, which correspond to $p_\mathrm{max}/E(B - V)$ = 9.0\% / mag \citep{Serkowski75}.
    The means, standard deviations, and average errors of $p/E(H - K_S)$ for the bulge sources are shown
    at the upper right of the panels (the means are also represented by arrows).
    The averages errors of $p/E(H - K_S)$ are derived from statistical and systematic errors of $p$ and $E(H - K_S)$.
    The dash-dotted lines in each panel show the means of $p/E(H - K_S)$ for the disk sources with $\delta p$ $\leq$ 1\%
    in each band.}
  \label{pehistogram}
\end{figure}

Combining means of $E(H - K_S)$ and $p$, we calculate the means $\langle$$p$$\rangle$/$\langle$$E(H - K_S)$$\rangle$
for the disk sources to be 16.2\% / mag, 9.2\% / mag, and 6.2\% / mag in the $J$, $H$, and $K_S$ bands, respectively
(dash-dotted lines in Fig. \ref{pehistogram}).
We show histograms of $p/E(H - K_S)$ for the bulge sources in Figure \ref{pehistogram}.
The means of $p/E(H - K_S)$ for the bulge sources are 6.3\% / mag, 3.7\% / mag, and 2.4\% / mag in the $J$, $H$,
and $K_S$ bands, respectively (shown as arrows in the histograms).
The average polarization efficiency of the ISM between the GC and us is considerably lower than that of the ISM in the disk,
by a factor of about three.

We make a comparison between observed polarization efficiency and estimated upper limits
[$p_J/E(H - K_S)$ = 25.0\% / mag, $p_H/E(H - K_S)$ = 14.5\% / mag, and $p_{K_S}/E(H - K_S)$ = 9.0\% / mag],
which are estimated by extending the upper limit [$p_\mathrm{max}/E(B - V)$ = 9.0\% / mag]
at optical wavelengths \citep{Serkowski75} to NIR wavelengths as follows.
We convert color excess $E(B - V)$ to $E(H - K_S)$ by assuming that $E(H - K)$ of the interstellar extinction law
by \citet{Rieke85} is identical to $E(H - K_S)$; $E(B - V)/E(H - K_S)$ = 5.14.
As for degree of polarization, first, $p_\mathrm{max}$ is converted to $p_K$ (the value at 2.2 $\micron$) using the average ratio%
$\langle$$p_\mathrm{max}/p_K$$\rangle$ = 5.4 \citep{Jones89,Wilking80}.
Then $p_{K_S}$ are extrapolated from $p_K$ following a power law $p_\lambda \propto \lambda^{-1.76}$,
and $p_J$ and $p_H$ are calculated from $p_{K_S}$ using $p_H/p_J$ = 0.581 and $p_{K_S}/p_H$ = 0.620
(\S\ref{THE WAVELENGTH DEPENDENCE OF POLARIZATION}).
The average polarization efficiency of the ISM in the disk is about two-thirds of the upper limits.
Moreover, that of the ISM between the GC and us is no more than about a quarter of the estimated upper limits.
\citet{Kobayashi83} also suggested that the polarization efficiency toward the GC is considerably lower than
that obtained in the solar neighborhood based on $K$ band polarimetry toward the GC (20$\arcmin$ $\times$ 20$\arcmin$).

\begin{figure}
  \epsscale{1.0}
  \plotone{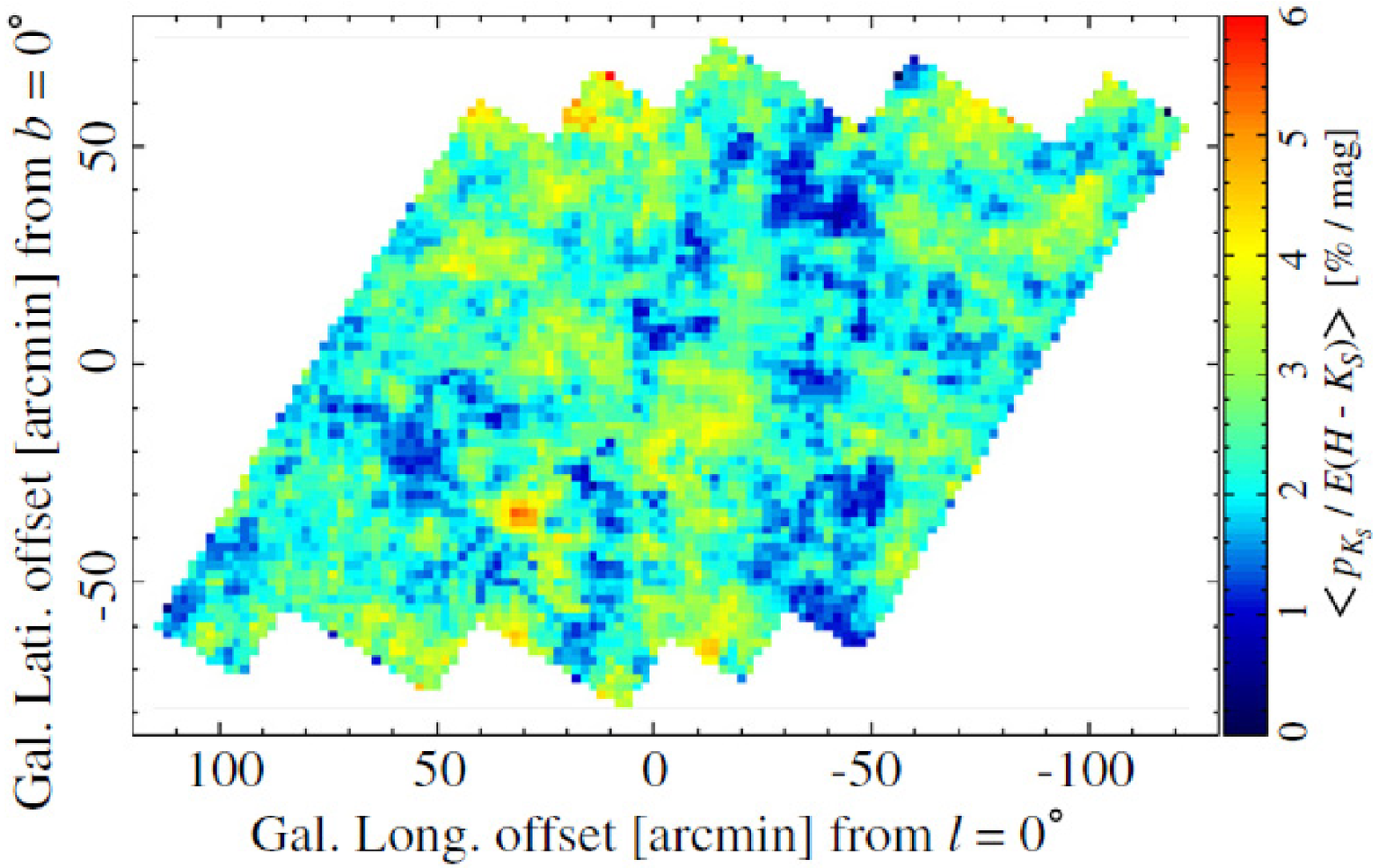}
  \caption{Map of $p_{K_S}/E(H - K_S)$.
    Each pixel represents a mean of $p_{K_S}/E(H - K_S)$ for the bulge sources with $\delta p_{K_S}$ $\leq$ 1\%
    in each cell with a size of 2$\arcmin$ $\times$ 2$\arcmin$.}
  \label{kpetile}
\end{figure}

The standard deviations of polarization efficiency $\sigma$($p/E(H - K_S)$), which are larger than the average errors of $p/E(H - K_S)$,
show that the polarization efficiency has spatial variation.
As for $E(H - K_S)$, $\theta_{K_S}$, and $p_{K_S}$,
we present a map of $p_{K_S}/E(H - K_S)$ for the bulge sources in Figure \ref{kpetile}.
It shows large variation from line-of-sight to line-of-sight in a range of about 1 to 5\% / mag,
but does not depend on the Galactic longitude and latitude, nor on the Galactic structure.

\begin{figure}
  \epsscale{1.0}
  \plotone{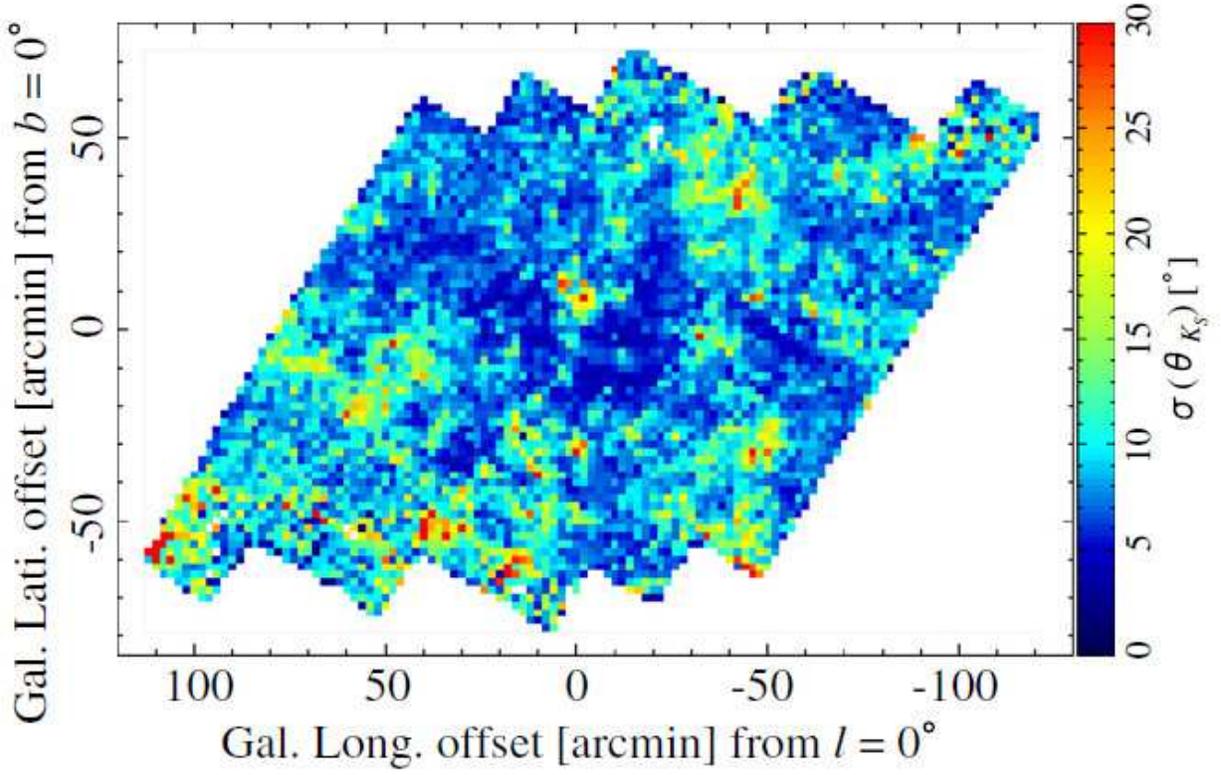}
  \caption{Map of $\sigma$($\theta_{K_S}$).
    Each pixel represents a dispersion of $\theta_{K_S}$ for the bulge sources with $\delta p_{K_S}$ $\leq$ 1\%
    and $\delta \theta_{K_S}$ $\leq$ 10$\arcdeg$ in each cell with a size of 2$\arcmin$ $\times$ 2$\arcmin$.
    The cells including one or no source(s) are drawn by white pixels.}
  \label{kpastddevtile}
\end{figure}

\begin{figure}
  \epsscale{1.0}
  \plotone{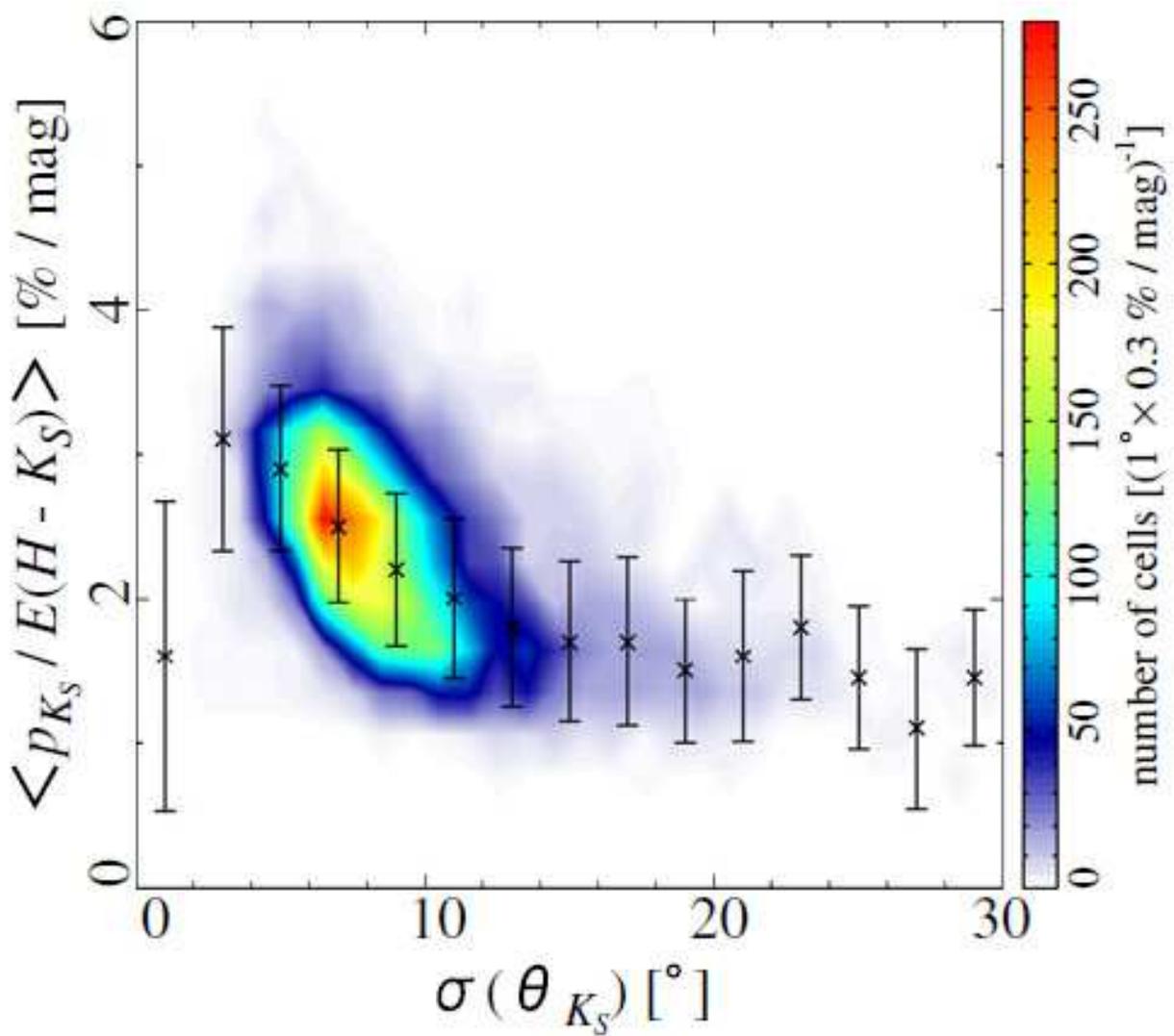}
  \caption{$\langle$$p_{K_S}/E(H - K_S)$$\rangle$ vs. $\sigma$($\theta_{K_S}$) for the cells including more than two sources.
    The crosses and error bars represent medians and standard deviations of $\langle$$p_{K_S}/E(H - K_S)$$\rangle$
    in 2$\arcdeg$ width bins of $\sigma$($\theta_{K_S}$).}
  \label{kpastddev_pe}
\end{figure}

Dispersions of position angles $\sigma$($\theta_{K_S}$) anticorrelate with $\langle$$p_{K_S}/E(H - K_S)$$\rangle$.
We calculate $\sigma$($\theta_{K_S}$) for the bulge sources with $\delta \theta_{K_S}$ $\leq$ 10$\arcdeg$ in each cell
and show its map in Figure \ref{kpastddevtile}.
Some cells show $\sigma$($\theta_{K_S}$) significantly larger than
the average error of $\theta_{K_S}$ for the bulge sources (5$\fdg$7).
In comparison between the spatial variations of $p_{K_S}/E(H - K_S)$ and $\sigma$($\theta_{K_S}$)
(Figs. \ref{kpetile} and \ref{kpastddevtile}),
we can see a tendency that the larger $\sigma$($\theta_{K_S}$), the lower $\langle$$p_{K_S}/E(H - K_S)$$\rangle$, and vice versa.
This tendency is shown in the relation between $\langle$$p_{K_S}/E(H - K_S)$$\rangle$ and $\sigma$($\theta_{K_S}$)
(Fig. \ref{kpastddev_pe}).
The medians of $\langle$$p_{K_S}/E(H - K_S)$$\rangle$ in each bin of $\sigma$($\theta_{K_S}$) (2$\arcdeg$ in width)
show the highest value at the bin of $\sigma$($\theta_{K_S}$) = 2$-$4$\arcdeg$,
decrease with increasing $\sigma$($\theta_{K_S}$), and then become almost flat toward bins of larger $\sigma$($\theta_{K_S}$).

\subsection{THE WAVELENGTH DEPENDENCE OF POLARIZATION}
\label{THE WAVELENGTH DEPENDENCE OF POLARIZATION}

\begin{figure}
  \epsscale{1.0}
  \plotone{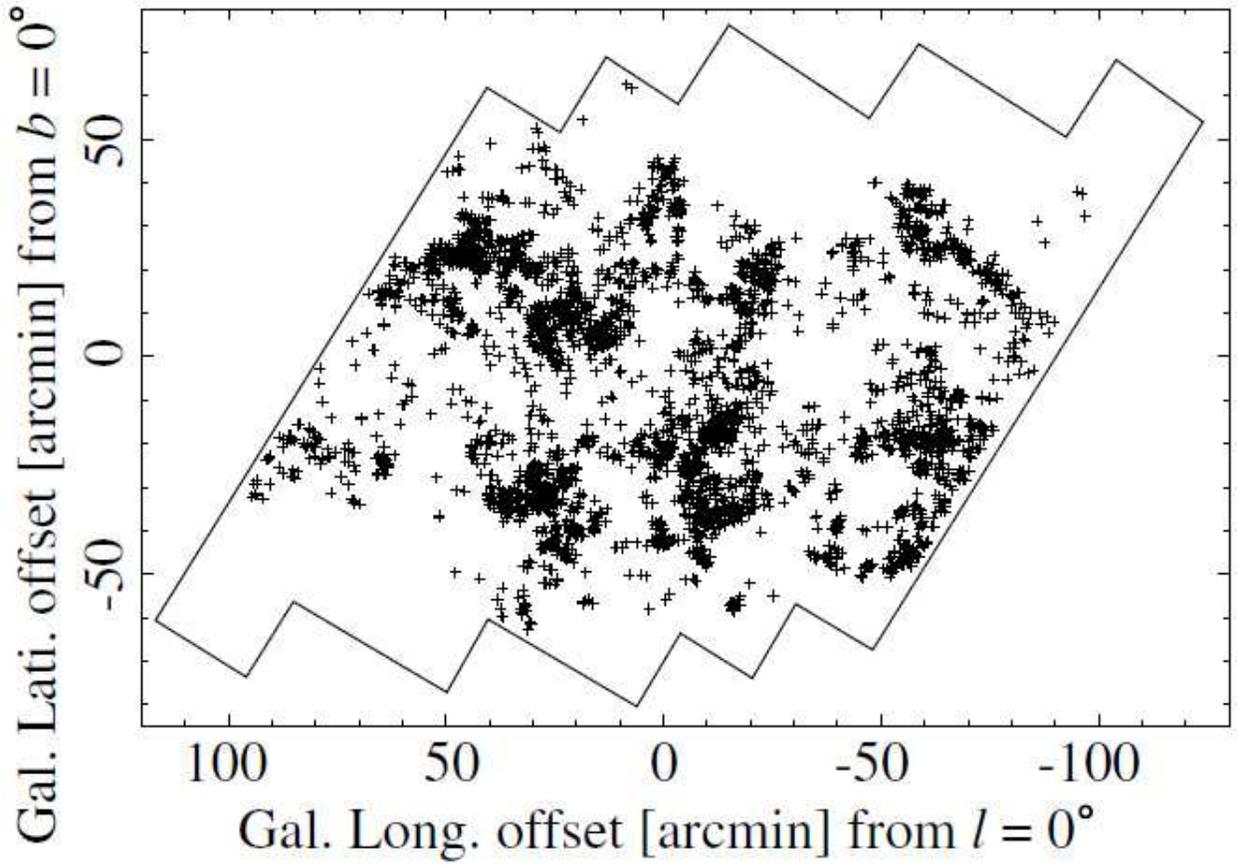}
  \caption{Spatial distribution of the sources that are detected in all the bands and have $\delta p$ $\leq$ 1\%
    and $p$ $\geq$ 10 $\delta p$ in all the bands. The solid lines show the observed area.}
  \label{sample}
\end{figure}

\begin{figure}
  \epsscale{0.6}
  \plotone{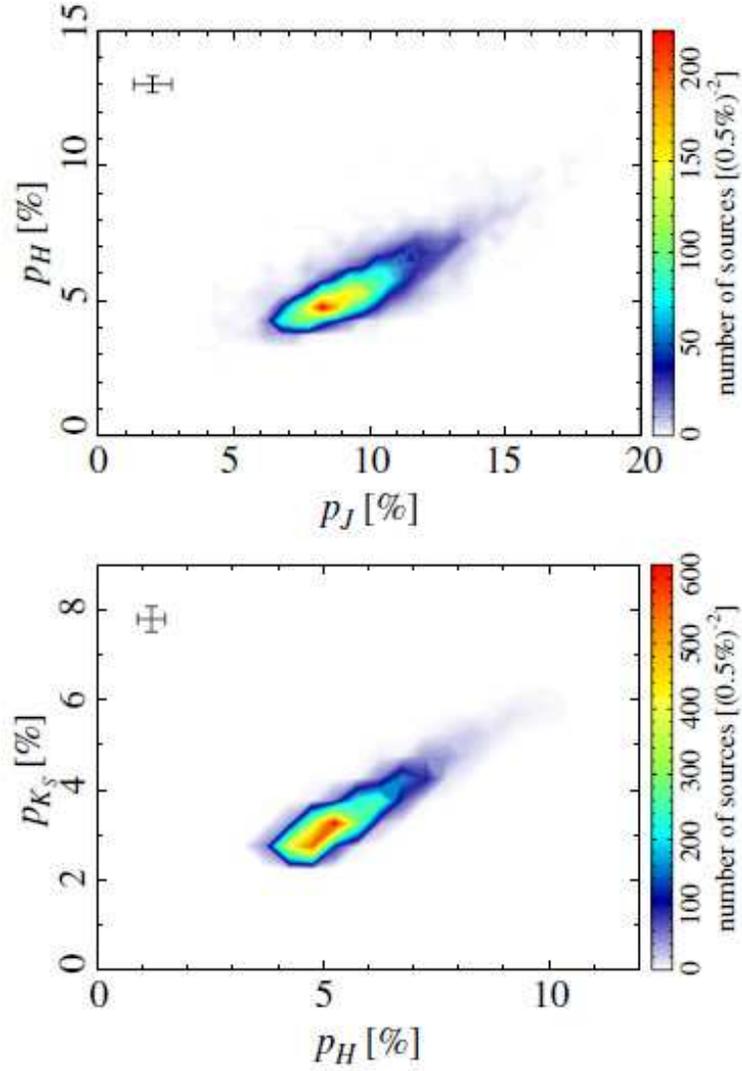}
  \caption{$p_J$ vs. $p_H$ ($top$) and $p_H$ vs. $p_{K_S}$ ($bottom$) for the sources that are detected in all the bands
    and have $\delta p$ $\leq$ 1\% and $p$ $\geq$ 10 $\delta p$ in all the bands.
    The upper left crosses in each panel denote the average errors of $p$.}
  \label{prelation}
\end{figure}

We here examine the wavelength dependence of polarization.
Out of the detected sources, we select 3,651 sources that have $\delta p$ $\leq$ 1\% and $p$ $\geq$ 10 $\delta p$ in all the bands.
Hereafter we use these sources in this section.
3,647 of these sources have $H - K_S$ $\geq$ 0.4 mag, and therefore they are bulge sources (\S\ref{MK CLASSIFICATIONS AND LOCATIONS}).
Figure \ref{sample} shows the spatial distribution of the sources.
The distribution of these bulge sources is little clumpy on regions where extinction is moderate (not especially small and large)
but the sources are seen in diverse area in $l$, $b$ (see also Fig. \ref{kexttile}).
In Figure \ref{prelation}, the correlations $p_J$ vs. $p_H$ and $p_H$ vs. $p_{K_S}$ are good with linear regressions of
$\langle$$p_H/p_J$$\rangle$ = 0.581 $\pm$ 0.004 and $\langle$$p_{K_S}/p_H$$\rangle$ = 0.620 $\pm$ 0.002 (Table \ref{dependence}).
The scatter around the best-fitting lines are 0.076 and 0.047 for $p_H/p_J$ and $p_{K_S}/p_H$, respectively.
The scatter can be explained by the observational error in $p_\lambda$:
the average errors are $\langle$$\delta(p_H/p_J)$$\rangle$ = 0.055 and $\langle$$\delta(p_{K_S}/p_H)$$\rangle$ = 0.063.
Thus the wavelength dependence of polarization does not change significantly from line-of-sight to line-of-sight in our sample.

\begin{figure}
  \epsscale{0.6}
  \plotone{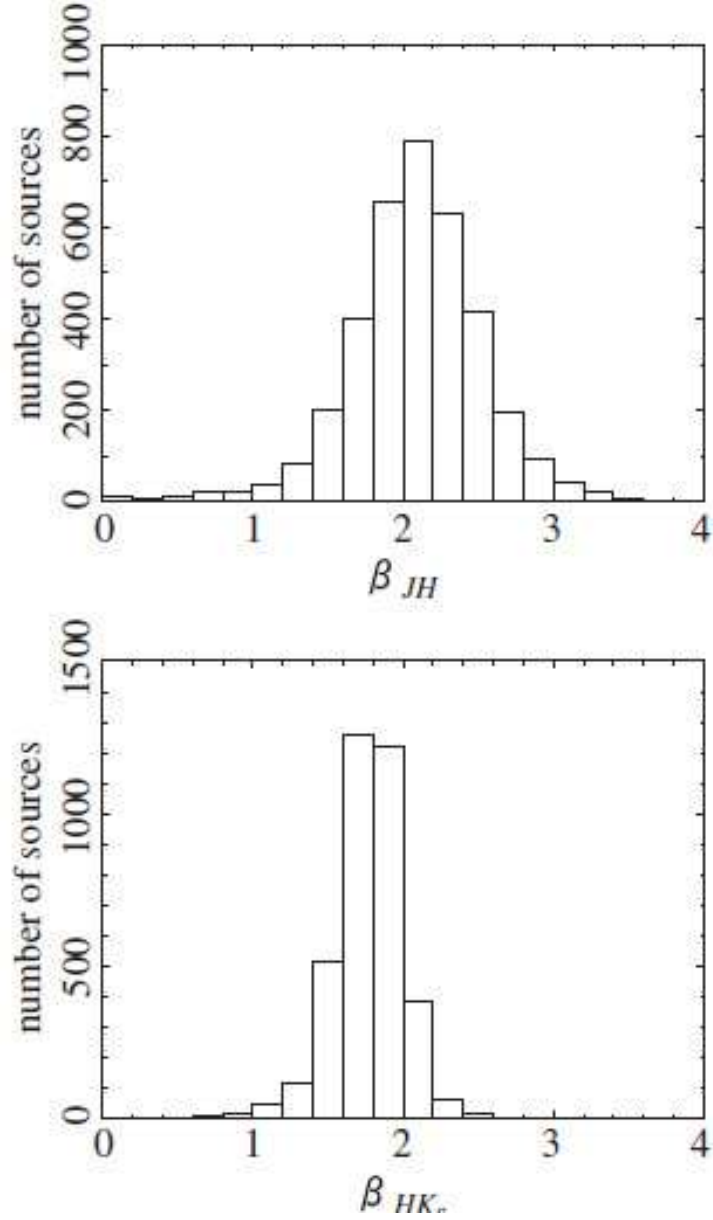}
  \caption{Histograms of $\beta_{JH}$ ($top$) and $\beta_{HK_S}$ ($bottom$) for the sources
    that are detected in all the bands and have $\delta p$ $\leq$ 1\% and $p$ $\geq$ 10 $\delta p$ in all the bands.
    The means, standard deviations, and average errors of $\beta_{JH}$ and $\beta_{HK_S}$
    for the sources are shown in Table \ref{dependence}.}
  \label{beta}
\end{figure}

Assuming a power law ($p_\lambda \propto \lambda^{-\beta}$), and using the equations:
\begin{displaymath}
  \beta_{JH} = -\frac{\mathrm{ln} (p_H/p_J)}{\mathrm{ln} (\lambda_H/\lambda_J)},
\end{displaymath}
\begin{displaymath}
  \beta_{HK_S} = -\frac{\mathrm{ln} (p_{K_S}/p_H)}{\mathrm{ln} (\lambda_{K_S}/\lambda_H)},
\end{displaymath}
we calculate indices $\beta_{JH}$ and $\beta_{HK_S}$ for the sources.
Histograms of $\beta_{JH}$ and $\beta_{HK_S}$ are shown in Figure \ref{beta}.
The means $\langle$$\beta_{JH}$$\rangle$ and $\langle$$\beta_{HK_S}$$\rangle$ for the sources are
2.08 $\pm$ 0.02 and 1.76 $\pm$ 0.01, respectively (Table \ref{dependence}).
The errors of the means and the average errors of $\beta_{JH}$ and $\beta_{HK_S}$ for the sources are derived
in a similar manner to $p_H/p_J$ and $p_{K_S}/p_H$.
Although these values are not inconsistent with the empirical values of 1.6$-$2.0 \citep{Nagata90,Martin90,Martin92},
$\langle$$\beta_{JH}$$\rangle$ is larger than $\langle$$\beta_{HK_S}$$\rangle$.
The degree of polarization decreases more slowly than a power law as the wavelength becomes longer from 1.25 to 2.14 $\micron$;
the wavelength dependence of polarization appears to flatten toward longer wavelengths.

\section{DISCUSSION}

\subsection{LOW POLARIZATION EFFICIENCY AND ITS SPATIAL VARIATION}

As for the polarization efficiency of the ISM between the GC and us, we revealed
(a) low efficiency compared to that of the ISM in the disk (Fig. \ref{pehistogram}),
(b) the spatial variation throughout the observed area (Fig. \ref{kpetile}),
and (c) the anticorrelation with the dispersions of position angles (Fig. \ref{kpastddev_pe}).
To explain our results, we discuss the polarization efficiency in relation to two factors:
(1) the polarizing grain properties and (2) the magnetic field direction.

The polarizing grain properties such as their shape, size distribution, and composition could affect the polarization efficiency.
However, the wavelength dependence of polarization shows little or no spatial variation in our sample
(\S\ref{THE WAVELENGTH DEPENDENCE OF POLARIZATION}).
This suggests that the polarizing grain properties are almost uniform,
and that the result of (b) cannot be explained by the factor (1), the polarizing grain properties.

Superposition of the ISM with different magnetic field directions along the line-of-sight could affect the polarization efficiency.
The lines-of-sight towards the disk and bulge sources crosses multiple ISM with a range of physical conditions.
The interstellar magnetic field consists of the uniform and random components \citep{Heiles87,Heiles96}.
The uniform component corresponds to the large-scale Galactic magnetic field,
which runs almost parallel to the spiral arms \citep{Heiles96,Han09}.
The magnetic field direction in a given ISM segment
(defined as a part of ISM; in each segment, both the magnetic field direction and the degree of grain alignment are constant)
can deviate from the direction of the Galactic magnetic field due to the presence of a random local component.
Due to the differences of the magnetic field directions, superposition of the ISM along the line-of-sight
lowers the polarization efficiency (depolarization).
Since the ISM between the GC and us generally consists of more ISM with different magnetic field directions,
depolarization would be larger and the polarization efficiency should be lower than the ISM in the disk (result of (a)).
To explain the observed dispersions of position angles (Fig. \ref{kpastddevtile}),
nonuniform structures of the magnetic field and/or density with a size of less than 2$\arcmin$ (cell size) are needed.
\citet{Gosling06,Gosling09} detected such nonuniform density distribution with a size of 5$-$15$\arcsec$.
Larger (smaller) differences of the magnetic field directions would cause lower (higher) polarization efficiency
and a larger (smaller) dispersion of position angles (results of (b) and (c)).
Of these two factors, only the factor (2), the differences of the magnetic field directions along the line-of-sight,
can explain our results.

\subsection{THE MAGNETIC FIELD STRENGTH RATIO OF THE RANDOM TO THE UNIFORM COMPONENT}

We discuss the magnetic field strength ratio of the random to the uniform component
based on the observed relation between extinction and degree of polarization.
For the relation, \citet{Jones92} constructed two models depending on the geometry of magnetic fields along the line-of-sight.
One is the two-component model, and the other is the wave model.

\begin{figure}
  \epsscale{0.7}
  \plotone{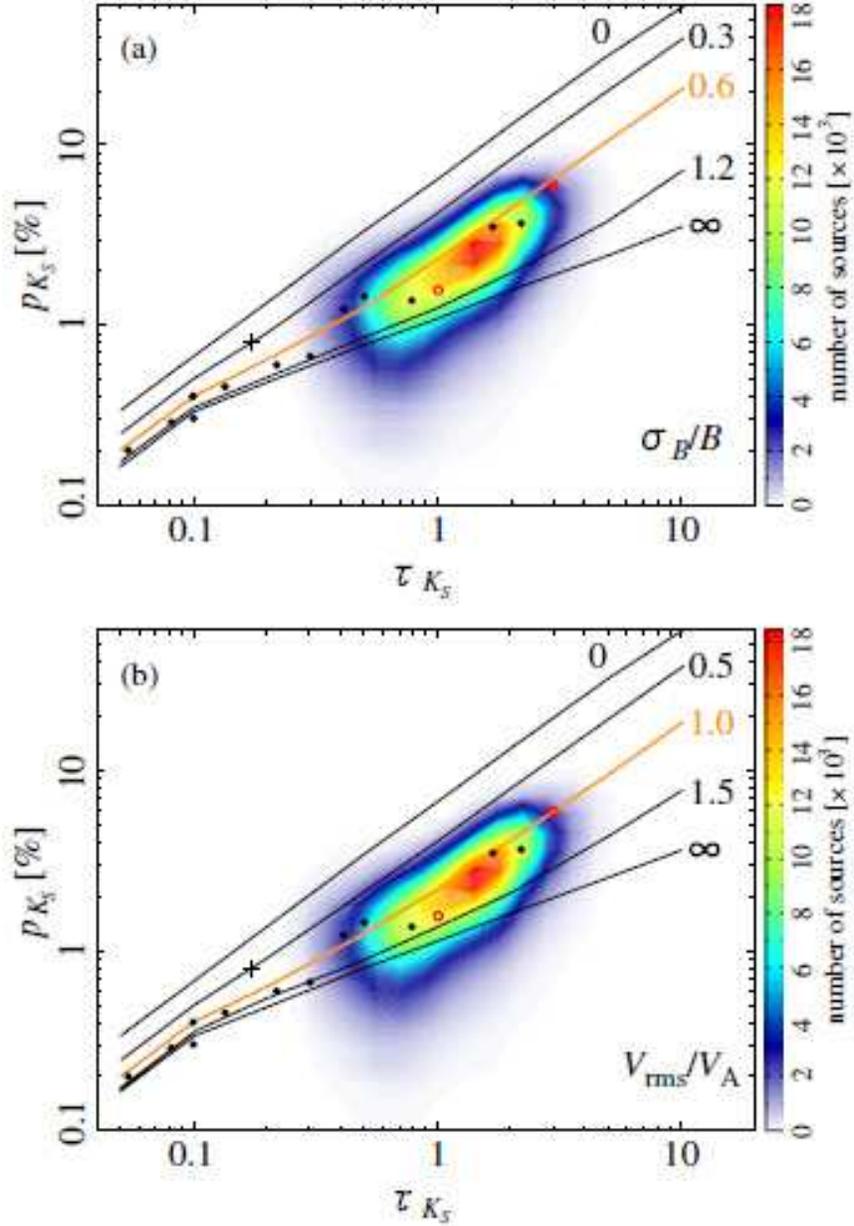}
  \caption{Degrees of polarization $p_{K_S}$ vs. optical depths $\tau_{K_S}$ for the bulge sources with $\delta p_{K_S}$ $\leq$ 1\%.
    The cross indicates the means of $p_{K_S}$ and $\tau_{K_S}$ for the disk sources with $\delta p_{K_S}$ $\leq$ 1\%.
    The lines show ($a$) the results for five values for the parameter
    $\sigma_\textbf{\textit{\scriptsize B}}/\textbf{\textit{B}}$ in the two-component model
    and ($b$) those for $V_\mathrm{rms}/V_\mathrm{A}$ in the wave model by \cite{Jones92}.
    The thick lines are their best-fit results.
    The circles show average polarization seen toward several regions at various optical depth intervals
    and the central regions of several normal spiral galaxies \citep[][and references therein]{Jones92},
    of which the open circles correspond to the measurements toward the GC by \citet{Kobayashi83}.}
  \label{kptau}
\end{figure}

In the two-component model, the magnetic field direction is determined by
a combination of the uniform and random components in each optical depth length (segment);
in each length $\Delta \tau_{K_S}$ = 0.1, the random component of the magnetic field decorrelates.
A segment corresponds to a part of the diffuse ISM (with a typical length of a few tens of pc) or a dense cloud (a fraction of a pc).
Fitting the model to the data, \citet{Jones92} concluded that the uniform and random components have equal energy density;
this is their case of $\sigma_\textbf{\textit{\scriptsize B}}/\textbf{\textit{B}}$ = 0.6,
where $\sigma_\textbf{\textit{\scriptsize B}}$ is the dispersion of the random component
and \textbf{\textit{B}} is the strength of the uniform component.
We compare our data with their results in the model (Fig. \ref{kptau}$a$).
Optical depths $\tau_{K_S}$ are calculated from color excess $E(H - K_S)$ using the relations
$A_{K_S}$/$E(H - K_S)$ = 1.44 \citep{Nishiyama06} and $\tau_{K_S}$ = $A_{K_S}$/2.5 log$_{10}$e.
Some of the bulge sources are distributed below the boundary with $\sigma_\textbf{\textit{\scriptsize B}}/\textbf{\textit{B}}$ = $\infty$.
Taking errors of $\delta p_{K_S}$ $\leq$ 1\% into consideration, these measurements can move to the region above the boundary.
The bulge sources show relatively lower polarization efficiency than their best-fit result
and almost lie between $\sigma_\textbf{\textit{\scriptsize B}}/\textbf{\textit{B}}$ = 0.6
and $\sigma_\textbf{\textit{\scriptsize B}}/\textbf{\textit{B}}$ = 1.2.
The polarization efficiency toward the GC measured by \citet{Kobayashi83} as shown by the open circles
is also lower than their best-fit result.
These indicate that the energy density of the random component is higher than that of the uniform component of
the magnetic field toward the GC.

In the wave model, a magnetic field is described as a wave.
The amplitude of the wave determines the extent to which the magnetic field direction in each segment fluctuates
along the line-of-sight.
\citet{Jones92} fitted the model to the data, also concluding that
the energy density of the magnetic field is in equipartition with the kinematic energy density of moving clouds;
this is their case of $V_\mathrm{rms}/V_\mathrm{A}$ = 1.0, where $V_\mathrm{rms}$ is the rms motion of individual clouds of
gas and dust attached to the magnetic field lines and $V_\mathrm{A}$ is the Alfv\'{e}n speed.
The comparison between our data and their results in the model is shown in Figure \ref{kptau}$b$.
The measurements below the boundary with $V_\mathrm{rms}/V_\mathrm{A}$ = $\infty$ can also move to the region above the boundary
if we take errors of $\delta p_{K_S}$ $\leq$ 1\%.
Compared to their best-fit result with $V_\mathrm{rms}/V_\mathrm{A}$ = 1.0,
the bulge sources show relatively lower polarization efficiency,
most of which are distributed between $V_\mathrm{rms}/V_\mathrm{A}$ = 1.0 and $V_\mathrm{rms}/V_\mathrm{A}$ = 1.5.
This means that the turbulent energy density is higher than the magnetic energy density in the ISM toward the GC.

The comparison in either case suggests a higher magnetic field strength of the random component compared to
that of the uniform component between the GC and us.
Such a trend is also observed in the solar neighborhood \citep{Heiles96,Fosalba02}.
The gas motions such as turbulence, gravitational contraction of dense clouds,
expansion of H {\footnotesize I\hspace{-.1em}I} regions, and supernova explosions distort the magnetic field
and produce its random component.
These processes would be strongly active in the direction of the GC \citep[see e.g., a review by][]{Morris96},
and contribute to the higher magnetic field strength ratio of the random to the uniform component toward the GC.
The processes cause relatively large deviations of the magnetic field directions from the direction of the uniform component
(the Galactic magnetic field).
Low polarization efficiency and its spatial variation would be explained by
superposition of the diffuse ISM and dense clouds with such deviations of the magnetic field directions along the line-of-sight.

\subsection{FLATTENING IN THE WAVELENGTH DEPENDENCE OF POLARIZATION}

In our results, $\langle$$\beta_{JH}$$\rangle$ is larger than $\langle$$\beta_{HK_S}$$\rangle$;
the wavelength dependence of polarization shows flattening from 1.25 to 2.14 $\micron$.
In previous studies \citep{Wilking80,Wilking82,Nagata90,Creese95}, similar flattening is also seen.

The value $\langle$$\beta_{JH}$$\rangle$ (2.08 $\pm$ 0.02) is close to
the power law index $\alpha$ (1.99 $\pm$ 0.02) of the wavelength dependence of \textit{extinction} toward the GC \citep{Nishiyama06}.
However $\langle$$\beta_{HK_S}$$\rangle$ (1.76 $\pm$ 0.01) is clearly below $\alpha$,
and flattening cannot be seen in the wavelength dependence of extinction from the $H$ to $K_S$ band.
At longer wavelengths beyond 3 $\micron$, both the wavelength dependence of polarization \citep{Nagata90,Martin92,Nagata94}
and extinction \citep[][and references therein]{Nishiyama09b} show flattening.

The wavelength dependence of polarization is determined by the polarizing grain properties such as shape, size distribution,
and composition.
From a comparison between observational data and theoretical models, the polarizing grain properties can be examined.
\citet{Kim94,Kim95a,Kim95b} fitted models of infinite cylindrical and spheroidal dust grains to modified Serkowski's law
\citep{Whittet92} and a single power law ($\beta$ = 1.65) for $\lambda$ = 1.64$-$5 $\micron$ to examine the size (mass)
distribution of dust grains.
They obtained the most satisfactory result by adopting perfectly aligned oblate dust grains (axial ratio of 6:1).
The resultant mass distribution has a peak at dust size of about 0.2 $\micron$
and a shoulder from the peak through dust size of 0.6 to 1.0 $\micron$ \citep[see Fig. 3$b$ of][]{Kim95b}.
The shoulder is required to fit the infrared polarization with a power law behavior,
which is the excess above the Serkowski's law.
To explain the flattening (i.e., the excess above a power law behavior),
greater numbers of such large-size dust grains would be necessary.

\section{CONCLUSION}

We have made polarimetric imaging observations toward the GC in order to examine the efficiency
and wavelength dependence of interstellar polarization at NIR.
The results are as follows.

1. The polarization efficiency of the ISM between the GC and us is lower than that of the ISM in the disk,
by a factor of about three on average.

2. The spatial variation of the polarization efficiency does not depend on the Galactic structure in contrast with
those of color excess and degree of polarization.

3. Position angles are almost parallel to the Galactic plane, suggesting that the magnetic field between the GC and us
has the longitudinal configuration and connects to the toroidal magnetic field in the GC.

4. The dispersions of position angles increase with decreasing the polarization efficiency.
It is likely that the polarization efficiency is reduced by
the different directions of magnetic fields along the line-of-sight (depolarization).

5. The comparison of our data with the models by \citet{Jones92} suggests that
the random component has a higher strength than the uniform component of the magnetic field.

6. The ratios of degree of polarization are $p_H/p_J$ = 0.581 $\pm$ 0.004 and $p_{K_S}/p_H$ = 0.620 $\pm$ 0.002,
which correspond to $\beta_{JH}$ = 2.08 $\pm$ 0.02 and $\beta_{HK_S}$ = 1.76 $\pm$ 0.01
for the power law indices of the wavelength dependence of polarization.
The degree of polarization is higher than that expected from a single power law toward longer wavelengths
(flattening from 1.25 to 2.14 $\micron$).
The flattening invokes greater numbers of aligned large-size dust grains in the mass distribution derived by \citet{Kim95b}.

\acknowledgments

We would like to thank the staff at the SAAO for their support during the observations.
The IRSF/SIRIUS project was initiated and supported by Nagoya University, the National Astronomical Observatory of Japan,
and the University of Tokyo in collaboration with the SAAO.
H. H. and S. N. are financially supported by a Research Fellowship of the Japan Society for the Promotion of Science
for Young Scientists.
This work was supported by KAKENHI, Grant-in-Aid for Scientific Research (A) 19204018 and (C) 21540240.
This publication makes use of data products from the Two Micron All Sky Survey,
which is a joint project of the University of Massachusetts and the Infrared Processing and Analysis Center/California Institute
of Technology, funded by the National Aeronautics and Space Administration and the National Science Foundation.

\clearpage

\begin{deluxetable}{lcccccc}
  \tabletypesize{\small}
  \tablewidth{450pt}
  \tablecolumns{7}
  \tablecaption{Polarimetry of R CrA No. 88 \label{standard}}
  \tablehead{\colhead{Study} & \colhead{$p_J$ [\%]} & \colhead{$\theta_J$ [$\arcdeg$]} & \colhead{$p_H$ [\%]} &
    \colhead{$\theta_H$ [$\arcdeg$]} & \colhead{$p_{K_S} [\%]$} & \colhead{$\theta_{K_S}$ [$\arcdeg$]}}
  \startdata
    this study        & 3.83 $\pm$ 0.17 & 90 $\pm$ 1 & 2.66 $\pm$ 0.18 & 91 $\pm$ 1 & 1.60 $\pm$ 0.21                  & 92 $\pm$ 3 \\
    \citet{Whittet92} & 3.87 $\pm$ 0.06 & 90 $\pm$ 1 & 2.73 $\pm$ 0.07 & 92 $\pm$ 1 & 1.69 $\pm$ 0.08\tablenotemark{a} & 95 $\pm$ 1 \\
  \enddata
  \tablenotetext{a}{Note that the observations by \citet{Whittet92} were made in a non-standard $K$ passband
    whose central wavelength is 2.04 $\micron$.
    The value at 2.14 $\micron$ was calculated by the power law extrapolation of the values measured at 1.64 and 2.04 $\micron$
    in the same manner as \citet{Gerakines95}.}
\end{deluxetable}

\begin{deluxetable}{lcccc}
  \tabletypesize{\small}
  \tablewidth{420pt}
  \tablecolumns{5}
  \tablecaption{Wavelength Dependence of Polarization \label{dependence}}
  \tablehead{\colhead{} & \colhead{$p_H/p_J$} & \colhead{$p_{K_S}/p_H$} & \colhead{$\beta_{JH}$} & \colhead{$\beta_{HK_S}$}}
  \startdata
    mean               & 0.581 $\pm$ 0.004 & 0.620 $\pm$ 0.002 & 2.08 $\pm$ 0.02 & 1.76 $\pm$ 0.01 \\
    standard deviation &             0.076 &             0.047 &            0.46 &            0.25 \\
    average error      &             0.055 &             0.063 &            0.36 &            0.37 \\
  \enddata
\end{deluxetable}

\clearpage

\clearpage

\clearpage

\clearpage

\clearpage

\clearpage

\clearpage

\clearpage

\clearpage

\clearpage

\clearpage

\clearpage

\clearpage

\clearpage

\clearpage

\clearpage

\clearpage

\clearpage

\end{document}